\documentclass[10pt,aps,pre,noshowpacs,twocolumn,superscriptaddress,nobibnotes,nofootinbib,longbibliography]{revtex4-1} 

\usepackage{amssymb}
\usepackage{graphicx}
\usepackage{amsmath,amsfonts}
\usepackage[colorlinks=true,allcolors=blue]{hyperref}
\usepackage[utf8]{inputenc}
\usepackage{xcolor}  
\usepackage[normalem]{ulem}
\usepackage[T1]{fontenc} 
\usepackage{multirow}
\usepackage{booktabs}
\usepackage{threeparttable}
\usepackage{tabularx}
\usepackage{chemfig}
\usepackage[version=4]{mhchem}

\definecolor{darkcyan}{rgb}{0.0, 0.4, 0.6}

\definecolor{dartmouthgreen}{rgb}{0.05, 0.5, 0.06}

\definecolor{pegahcommentcolor}{rgb}{0.13, 0.3, 0.99}

\usepackage{pdfpages}
\makeatletter
\AtBeginDocument{\let\LS@rot\@undefined}
\makeatother

\begin{document}
    
    \title{Mapping bipartite networks into multidimensional hyperbolic spaces}
    
    \date{\today}
    
    \author{Robert Jankowski}
    \affiliation{Departament de F\'isica de la Mat\`eria Condensada, Universitat de Barcelona, Mart\'i i Franqu\`es 1, E-08028 Barcelona, Spain}
    \affiliation{Universitat de Barcelona Institute of Complex Systems (UBICS), Universitat de Barcelona, Barcelona, Spain}
    \author{Roya Aliakbarisani}
    \affiliation{Departament de F\'isica de la Mat\`eria Condensada, Universitat de Barcelona, Mart\'i i Franqu\`es 1, E-08028 Barcelona, Spain}
    \affiliation{Universitat de Barcelona Institute of Complex Systems (UBICS), Universitat de Barcelona, Barcelona, Spain}
    \author{M. {\'A}ngeles Serrano}
    \affiliation{Departament de F\'isica de la Mat\`eria Condensada, Universitat de Barcelona, Mart\'i i Franqu\`es 1, E-08028 Barcelona, Spain}
    \affiliation{Universitat de Barcelona Institute of Complex Systems (UBICS), Universitat de Barcelona, Barcelona, Spain}
    \affiliation{ICREA, Passeig Llu\'is Companys 23, E-08010 Barcelona, Spain}
    \author{Mari\'an Bogu\~{n}\'a}
    \email[]{marian.boguna@ub.edu}
    \affiliation{Departament de F\'isica de la Mat\`eria Condensada, Universitat de Barcelona, Mart\'i i Franqu\`es 1, E-08028 Barcelona, Spain}
    \affiliation{Universitat de Barcelona Institute of Complex Systems (UBICS), Universitat de Barcelona, Barcelona, Spain}

    \begin{abstract}
        Bipartite networks appear in many real-world contexts, linking entities across two distinct sets. One-mode projections, though widely used, can introduce artificial correlations and inflated clustering, obscuring the true underlying structure. In this work, we propose a geometric model for bipartite networks that leverages the high levels of bipartite four-cycles as a measure of clustering and embeds both node types into a shared similarity space, where the link probability decreases with distance. Additionally, we introduce B-Mercator, an algorithm that infers node positions from the bipartite structure. We evaluate its performance on diverse datasets, illustrating how the resulting embeddings improve downstream tasks such as node classification and distance-based link prediction. These hyperbolic embeddings enable realistic synthetic network generation with node features mirroring real data. By preserving the bipartite structure, our approach avoids projection biases, offering more accurate structural descriptions and providing a robust framework for uncovering hidden geometry in bipartite networks.
    \end{abstract}

    \maketitle
    \let\oldaddcontentsline\addcontentsline
    \renewcommand{\addcontentsline}[3]{}

    
    Bipartite networks lie at the heart of countless real-world applications, linking authors to the articles they write~\cite{newman2001structure,newman2001scientific,newman2001scientificII,Newman:2004rr}, users to the products they consume~\cite{koren2009matrix,konstan1997grouplens,linden2003amazon}, people to the groups they belong to~\cite{davis2009deep}, or countries to the languages they speak~\cite{unicode_cldr}. They also arise naturally in metabolic networks, where metabolites are connected to the chemical reactions or enzymes that transform them~\cite{jeong2000large,ravasz2002hierarchical,serrano2012uncovering}, in plant-pollinator networks~\cite{bascompte2003nested,olesen2007modularity}, and in machine learning applications, where nodes have associated features used to feed graph neural networks~\cite{aliakbarisani2023feature}. By definition, each bipartite system splits its nodes into two disjoint sets, with no edges connecting nodes within the same set. This seemingly simple rule nonetheless yields rich and complex connectivity patterns, enabling researchers to model collaboration, consumption, and association processes across diverse domains. Yet bipartite networks have historically garnered less attention than their unipartite counterparts. Given their ubiquity and straightforward interpretability, a renewed focus on bipartite structures 
    is both timely and necessary to fully capture the multiple facets of interaction present in many real-world complex systems and data structures. 
    
    A general practice is to analyze bipartite networks by projecting them onto a single node set,  
    creating a \emph{one-mode} network~\cite{zhou2007bipartite}. For instance, in an author--article bipartite network,  
    one might create a unipartite graph of authors by connecting two authors if they have co-authored at least one article~\cite{newman2001structure,newman2001scientific,newman2001scientificII,Newman:2004rr}. While such one-mode projections allow researchers to employ classic unipartite tools (e.g., clustering coefficients, degree distributions, community-detection algorithms), they also introduce strong correlations between edges. These correlations arise owing to sets of nodes that share common neighbors in the bipartite structure inducing cliques --or fully connected subgraphs-- in the one-mode projection. Hence, the resulting unipartite networks can exhibit inflated clustering and misleading connectivity patterns that do not necessarily reflect the independent pairwise interactions of the underlying bipartite system. Besides, there is an unavoidable loss of information when bipartite networks are projected and it is even possible to obtain the same one-mode projection out of different bipartite networks~\cite{zhou2007bipartite,neal2014backbone}.
    
    To overcome these limitations and more accurately capture the true structure of bipartite systems, it is critical to develop a modeling framework that treats bipartite networks directly rather than relying solely on their one-mode projections. 
    Thus, we turn our attention to the latent space models, initially developed for social networks~\cite{hoff2002latent} and extended to temporal bipartite networks~\cite{friel2016interlocking} and to growing bipartite generative models~\cite{oneale2020pso}. Within this family of models, we propose to use network geometry to make sense of bipartite networks and to find hyperbolic geometric representations for this class of systems~\cite{Krioukov2010}. Hyperbolic geometry is particularly useful for capturing the hierarchical organization of complex networks~\cite{Ortiz:2020fu}. In recent years, network geometry~\cite{Boguna2021} has been extremely successful in explaining undirected~\cite{Serrano2008}, directed~\cite{Allard:2024zh}, weighted~\cite{allard2017geometric}, and multiplex networks~\cite{kleineberg2016hidden,Kleineberg2017} in many real systems and has also been extended to bipartite settings, proposing that nodes of both types can indeed lie in a shared latent space with connection probabilities governed by their mutual distances~\cite{serrano2012uncovering,kitsak2017latent}. Despite the fact that bipartite networks do not contain triangles by definition (the signature of any metric space), it is still possible to define a clustering coefficient by counting cycles of length four (i.e., squares). Empirical analyses have shown that these four-cycles can be abundant in real bipartite networks, leading to high effective clustering values~\cite{OPSAHL2013159,aliakbarisani2023feature}. This finding suggests that, indeed, bipartite networks may be embedded in a metric space where the likelihood of a connection between two nodes of different types decreases with the distance separating them.
    
    Assuming that a bipartite network can be embedded in a similarity (or metric) space, a logical step is to seek methods for inferring the positions of its nodes within this space from real data. Previous approaches typically rely on one-mode projections of bipartite networks --using, for instance, the D-Mercator tool~\cite{Garcia2019,jankowski2023dmercator}-- to embed each type of node. However, such projections often introduce artificial correlations and inflated clustering, reducing embedding accuracy. Embeddings of one-mode projections have been used in~\cite{serrano2012uncovering} to infer the coordinates of one node type, which are then used to estimate the coordinates of the other type. Another approach~\cite{aliakbarisani2023feature} transfers coordinates from a unipartite layer --when such a layer exists-- and assumes that the same coordinates apply in the bipartite representation. However, none of these methods provides a model-based bipartite embedding. Finally,~\cite{kitsak2017latent} uses the number of common neighbors as a proxy for similarity distance between nodes, but these estimates are noisy and do not support a full embedding of the network.
    
    Here, following results in~\cite{serrano2012uncovering,kitsak2017latent}, we introduce the bipartite-$\mathbb{S}^D/\mathbb{H}^{D+1}$ model in which nodes of both types lie in the same $D$-dimensional similarity space. Using this model, we propose B-Mercator, an algorithm designed specifically for bipartite networks that enables the creation of multidimensional hyperbolic maps of real bipartite datasets. It is important to emphasize that choosing an appropriate similarity-space dimension $D$ is crucial: while many networks are well captured with $D=1$, others require higher $D$ to prevent distinct communities from being artificially mixed~\cite{Almagro2022,jankowski2023dmercator}. To illustrate its capabilities, we embedded and analyzed three datasets: Unicodelang~\cite{unicode_cldr}, which links countries to the languages spoken in those regions, Metabolic~\cite{king2016bigg}, in which metabolites are connected through chemical reactions, and Flavor~\cite{ahn2011flavor}, which connects food ingredients to their corresponding flavor compounds. Furthermore, we show how B-Mercator can be applied to supervised machine learning tasks, including node classification and link prediction, yielding a significant performance gain with respect to state of the art methods, especially when a strong correlation exists between nodes' labels and their feature distributions.

    \section{Results}
    
    \subsection{The bipartite-$\mathbb{S}^D/\mathbb{H}^{D+1}$ model and B-Mercator}
    
     \begin{figure}[t!]
        \centering
        \includegraphics[width=0.48\textwidth]{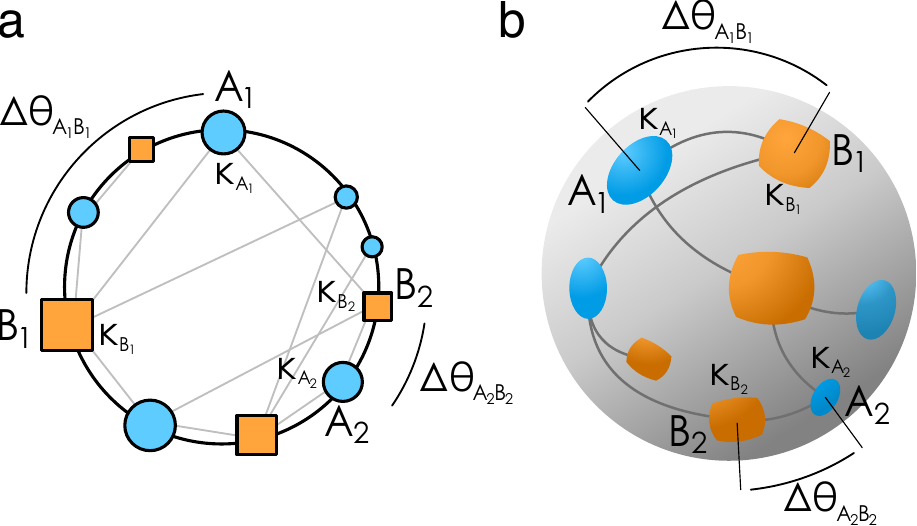}
        \caption{\textbf{Schematic representation of the bipartite-$\mathbb{S}^D$ model in dimension (a) $D=1$ and (b) $D=2$}. Nodes A are shown as circles whereas nodes B are shown as squares whose sizes are proportional to the nodes' expected degrees. The angular distances between nodes A and B are highlighted ($\Delta\theta_{A_1, B_1}$ and $\Delta\theta_{A_2, B_2}$). Light grey lines represent the edges in the bipartite network generated by Eq.~\eqref{eq:prob}.}
        \label{fig:schematic}
    \end{figure}

    The bipartite-$\mathbb{S}^D$ model generates bipartite networks with type-A and type-B nodes by assigning them coordinates on the surface of a $D$-sphere and expected degrees. Pairs of type-A and type-B nodes are then connected with a probability that depends on their spherical distance, rescaled by the product of their expected degrees. Figure~\ref{fig:schematic} shows two examples generated by the bipartite-$\mathbb{S}^D$ model for $D=1$ and $D=2$, where node size represents expected degree. The model has a purely geometric counterpart, the bipartite-$\mathbb{H}^{D+1}$ model, in which expected degrees are mapped to a radial coordinate so that nodes are placed in the $(D+1)$-sphere in the hyperbolic space and the connection probability becomes a function of the hyperbolic distance only. Thus, in the hyperbolic representation, angular coordinates on the $D$-sphere encode similarity, whereas the radial coordinate encodes popularity as reflected by the expected degree. In both representations, the model includes an inverse temperature $\beta_b$ that modulates noise and therefore controls the coupling between network topology and geometry.
        
     Given a real bipartite network, we treat it as a likely realization of the bipartite-$\mathbb{S}^D/\mathbb{H}^{D+1}$ model. B-Mercator then finds an embedding of both node types on the surface of a $D$-sphere $\mathbb{S}^D$ and their expected degrees (or, equivalently, in $\mathbb{H}^{D+1}$) that maximizes the likelihood under the bipartite-$\mathbb{S}^D/\mathbb{H}^{D+1}$ model. A full description of the model and the technical details of the B-Mercator embedding algorithm are provided in Methods, Sections~\ref{bipartiteS1H2} and~\ref{BMercator}.

    \subsection{Validation}
\begin{figure}[t!]
        \centering
        \includegraphics[width=0.45\textwidth]{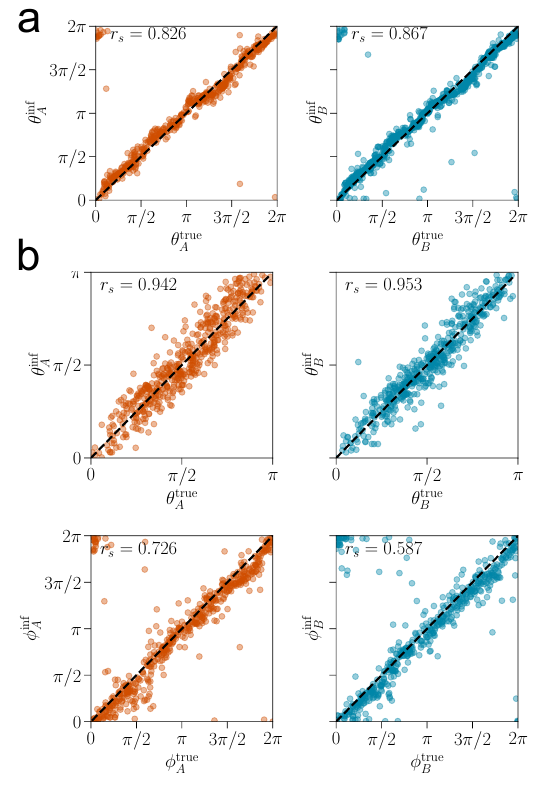}
        \caption{\textbf{Validation of B-Mercator on synthetic bipartite networks}. Relationship between the original and the inferred coordinates of the (\textbf{a}) bipartite-$\mathbb{S}^1$ and (\textbf{b}) bipartite-$\mathbb{S}^2$ models. In the top left corner of each figure, we report the value of the Spearman correlation coefficient between the inferred and original coordinates. Since the inferred coordinates might be rotated, we transform them to minimize the average angular distance between the original and inferred coordinates (Supplementary Section IV in \cite{jankowski2023dmercator}). 
            With parameters: $N_A=500$ (number of type-A nodes), $N_B=1000$ (number of type-B nodes), $\gamma_A=2.7$ (exponent of the powerlaw degree distribution of type-A nodes), $\gamma_B=2.1$ (exponent of the powerlaw degree distribution of type-B nodes), $\langle k_A \rangle=10$ (average degree of type-A nodes), $\beta_b = 1.5D$ (inverse temperature), with dimension $D=1$ for (\textbf{a}) and $D=2$ for (\textbf{b}).}
        \label{fig:validation}
    \end{figure}

    To validate our method, we generated single realizations of synthetic bipartite networks from the bipartite-$\mathbb{S}^D/\mathbb{H}^{D+1}$ model with different dimensions and parameters. We then used B-Mercator for each such synthetic network to infer both the model parameters and node coordinates $(\kappa^{\text{inf}},\theta^{\text{inf}})$, and compared them with the true values $(\kappa^{\text{true}},\theta^{\text{true}})$. Figure~\ref{fig:validation} shows comparisons between the true and inferred coordinates for type-A and -B nodes. One can observe a high Spearman correlation coefficient for $D=1$ and $D=2$, corroborating the effectiveness of our embedding technique. For more examples in $D=1$ and $D=2$ as well as $D=3$, see Supplementary Figures~2-4. B-Mercator can also infer the inverse temperature $\beta_b$ as shown in Supplementary Figure 7. 
    
    As a further check, we tested the reproducibility of the topological properties of the original network embedded by B-Mercator. Using the model parameters and coordinates inferred by B-Mercator, we generated an ensemble of synthetic networks using the bipartite-$\mathbb{S}^D/\mathbb{H}^{D+1}$ model (see Supplementary Figures 8--12). We then evaluated ensemble averages and standard deviations for various topological properties and compared them with those of the original network. The degree distributions and clustering spectra of type-A and type-B nodes were very well reproduced. We also observed good agreement between the empirical and theoretical connection probabilities. These findings confirm that B-Mercator accurately reconstructs the nodes' coordinates in synthetic networks and reliably determines other model parameters, such as hidden degrees and inverse temperature, regardless of the network dimensionality.

    \subsection{Bipartite greedy routing}
    
    In order to establish a meaningful geometric representation of a bipartite network at the global scale, we introduce bipartite greedy routing (BGR) as a practical tool to infer the network's effective dimension~\cite{boguna2009navigability}. The idea is to first embed the bipartite network into a latent geometric space with B-Mercator and then test how well nodes can route information by simply forwarding messages to their neighbors closest to the destination in that space. By systematically evaluating the success of these greedy routes --measured, for instance, by the probability that messages reach their targets without getting stuck-- we gain insights into the dimensional structure underlying the network. In this way, the performance of greedy routing serves as an indicator of how well the bipartite graph can be embedded in a space of a given dimension, effectively allowing us to determine the dimension that best captures its structure.

    \begin{figure}[h!]
        \centering
        \includegraphics[width=0.45\textwidth]{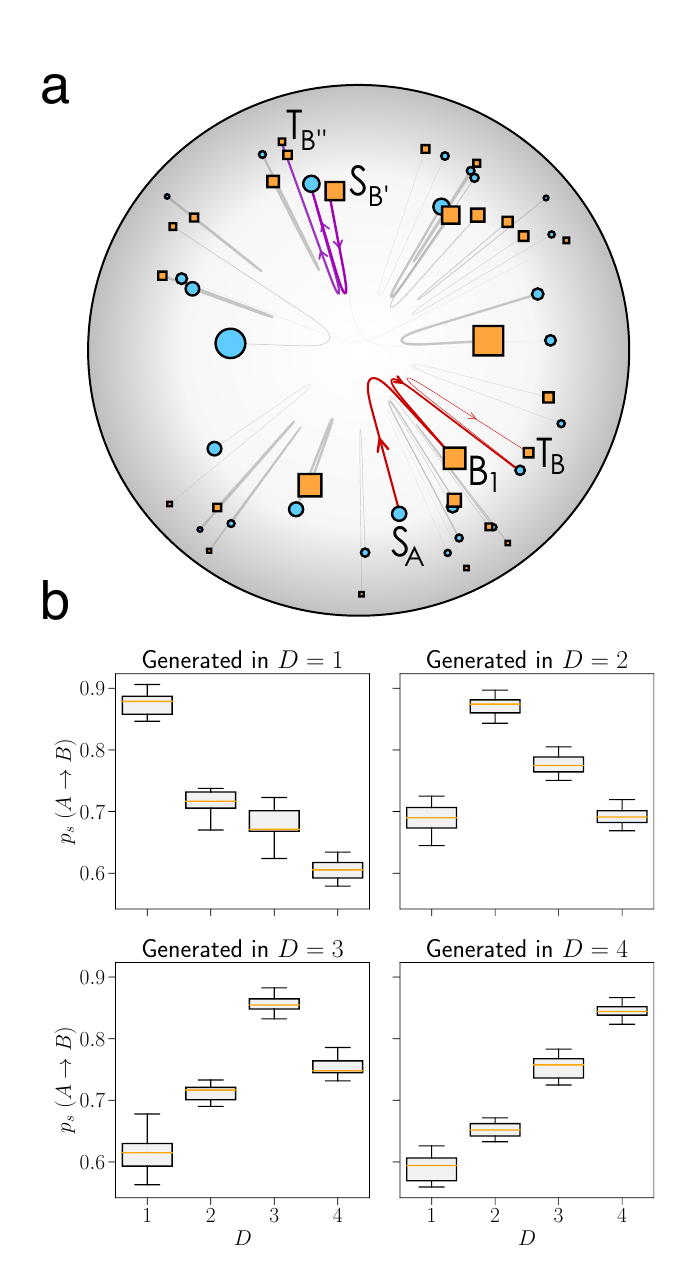}
        \caption{\textbf{Bipartite greedy routing in synthetic networks}. (a) Schematic view of the greedy routing protocol. We select a type-A node as an origin ($\mathbf{S_A}$) and a type-B ($\mathbf{T_B}$) as a destination. The red arrows show how the message is forwarded towards the destination. In the second example, we select two type-B nodes as source ($\mathbf{S_{B^\prime}}$) and destination ($\mathbf{T_{B^{\prime\prime}}}$)  and outline the greedy path with purple color. The line width is proportional to the connection probability (Eq.~\ref{eq:prob}). (b) Success rate as a function of embedded dimension for networks generated in $D=\{1,2,3,4\}$. We consider here a navigation protocol where a source is a type-A node and a destination a type-B one. Results are obtained by averaging over 10 realizations with parameters: $N_A=500$ (number of type-A nodes), $N_B=500$ (number of type-B nodes), $\gamma_A=2.5$ (exponent of the powerlaw degree distribution of type-A nodes), $\gamma_B=2.5$ (exponent of the powerlaw degree distribution of type-B nodes), $\langle k_A \rangle=10$ (average degree of type-A nodes), $\beta_b = 1.5D$ (inverse temperature). The box ranges from the first quartile to the third quartile. A horizontal line goes through the box at the median. The whiskers go from each quartile to the minimum or maximum.}
        \label{fig:bgr}
    \end{figure}
    
    We implemented a BGR protocol in which both the origin and destination can be either type-A or type-B nodes thus defining four variants. In Figure~\ref{fig:bgr}a, we depict a schematic picture of the BGR for the A--B variant, i.e., when the source node is a type-A node ($S_A$), and the target node is a type-B node ($T_B$). 
    The message is forwarded from $S_A$ to the type-B node that is hyperbolically closest to the target. 
    Since the node $B_1$ is not the destination, the message is forwarded again to a type-A node. The process is repeated until the destination is reached or the message becomes stuck. Then, the BGR protocol is executed for a large number of randomly chosen node pairs to assess the global network's geometric properties.
    
    We tested the BGR protocol in synthetic networks generated from the bipartite-$\mathbb{S}^D$ model. First, we generated networks with specific dimensionality and topological properties, and we obtained their hyperbolic maps by embedding them using B-Mercator with different embedding dimensions. The performance of BGR was assessed based on two key measures: the proportion of messages that successfully reach their destination $p_s$, i.e., the success rate, and the mean stretch, where stretch is defined for each path connecting a source and target node as the ratio of the hop count of a successful greedy path to the shortest path. In Figure~\ref{fig:bgr}b, we show the success rate as a function of the embedding dimension for networks generated using the bipartite-$\mathbb{S}^D$ model with dimensions ranging from $D=1$ to $D=4$. The performance of BGR is optimal (in terms of $p_s$ and mean stretch, see Supplementary Figure 14) when B-Mercator is used with the same dimension that was used to generate the network, thus justifying BGR as an alternative method to infer the effective dimension of real networks, different from the topological-based method introduced in~\cite{Almagro2022}. These results are consistent across all variants of BGR (see Supplementary Figure 13) and corroborate the results obtained for unipartite networks~\cite{jankowski2023dmercator}.

    \subsection{Embedding of real bipartite networks}
    
    The significance of B-Mercator lies not only in its ability to embed synthetic networks generated by the bipartite-$\mathbb{S}^D/\mathbb{H}^{D+1}$ model, but rather in its capacity to uncover geometric insights from real bipartite networks. Moreover, embeddings produced by B-Mercator can be applied to tasks such as node classification and link prediction on graph-structured data. As case studies, we analyze the Unicodelang dataset, which captures relationships between countries and the languages spoken within them, the human metabolic network connecting metabolites with the reactions in which they participate, and the flavor network linking ingredients to the chemical compounds they contain. These examples demonstrate the practical applicability of B-Mercator in extracting meaningful structural patterns from real-world bipartite networks.

    \begin{figure*}[t!]
        \centering
        \includegraphics[width=0.95\textwidth]{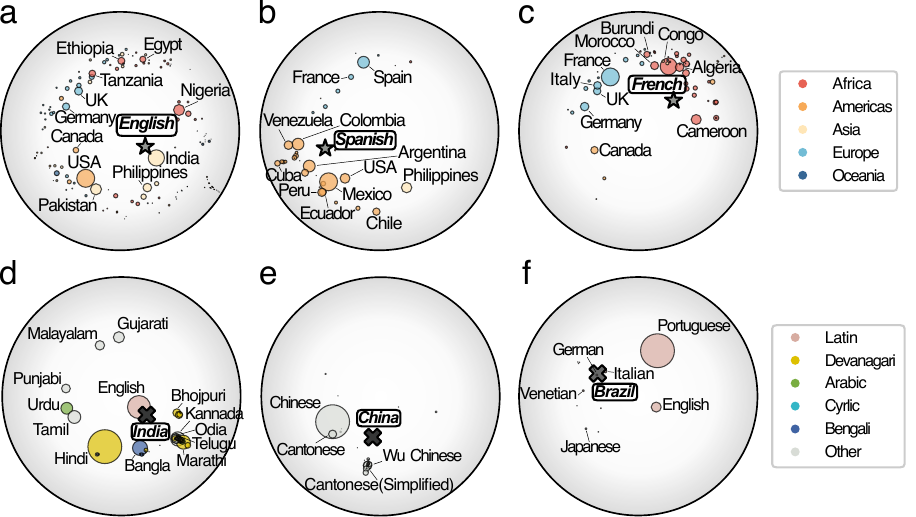}
        \caption{\textbf{Visualization of the bipartite-$\mathbb{S}^1$ embedding of the Unicodelang dataset per country or language}. Panels (\textbf{a, b, c}) show countries where a given language is spoken, i.e., the neighbors of the language node. The size of the nodes is proportional to the number of language speakers in that country. The color corresponds to the geographical region in which the country is located. A star marker indicates the position of a given language. In panels (\textbf{d, e, f}), we depict all languages spoken in a given country, i.e., the neighbors of the country node. The size of the nodes is proportional to the fraction of speakers of a given language. The color represents that language's script. A cross marker indicates the position of a given country.}
        \label{fig:lang}
    \end{figure*}

    Using B-Mercator, we embedded the Unicodelang dataset in various dimensions. The inferred embeddings are able to reproduce the topological properties of the network (see Supplementary Figure~17). In Figure~\ref{fig:lang}, we show a dual embedding representation of this dataset in dimension $D=1$. First, we focus on the countries in which a given language is used. For instance, in Figure~\ref{fig:lang}a, we plot all countries in which English is spoken, i.e., the neighbors of the English language in the bipartite network. One can notice that English is located closer to India or the Philippines than to the United States of America (USA). This can be explained by the fact that many different languages are spoken in the USA, which influence its position in the bipartite map. Indeed, in Figure~\ref{fig:lang}b, where we plot all neighbors of the Spanish language, the USA is located close to the Spanish language. Lastly, in Figure~\ref{fig:lang}c one can observe that countries from the French colonialism are concentrated in the similarity space and lie close to the French language, in contrast to France and other European countries. 
    We can also shift the perspective: instead of examining the neighbors of each language, we can analyze the neighbors of a given country. In the bottom row of Figure~\ref{fig:lang}, we plot the language neighbors of India, China, and Brazil. Countries are often located in the embedding space close to the most widely spoken language. 
    Additional examples are provided in Supplementary Figure~15, where we explore the neighbors of languages such as Hindi, Swahili, Catalan, Persian, Korean, Dutch, Russian, and Arabic. Similarly, in Supplementary Figure~16, we depict the neighbors of countries including Canada, Turkey, Indonesia, Cameroon, Tanzania, Greece, the Philippines, and Bolivia.

    \begin{figure}[t!]
        \centering
        \includegraphics[width=0.47\textwidth]{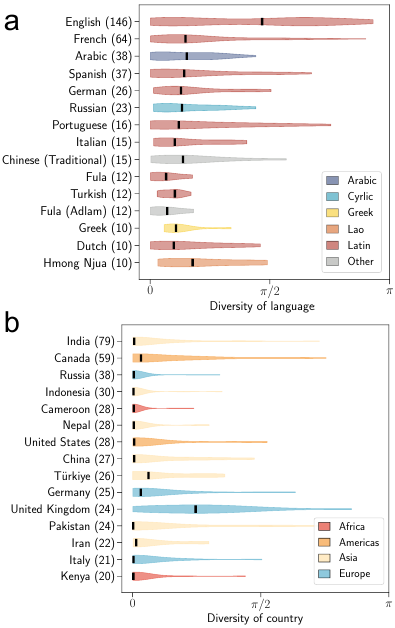}
        \caption{\textbf{Language or country diversity for the top 15 highest-degree nodes}. Panel (a) shows violin plots of angular distances between each language and its neighboring countries, with colors indicating script type. Panel (b) presents analogous plots for countries, with colors representing geographic region. In both panels, nodes are ordered in descending order of degree, with each node’s degree (in brackets) indicated next to its label. Only the central 95\% of the data is plotted--that is, data between the 2.5th and 97.5th quantiles are shown. A black line highlights the median value in each plot.}
        \label{fig:unicodelang_profile}
    \end{figure}

    The hyperbolic bipartite embedding enables us to examine the concentration of countries sharing the same language. We select the top 15 languages with the highest degree and compute the angular distance from each language to its neighboring countries. A small average angular distance may indicate that these countries are highly similar, whereas a broad distribution suggests a more international language. Figure~\ref{fig:unicodelang_profile}a displays the \textit{diversity} for these high-degree languages. As expected, the angular distance distribution for English is broad, indicating connections with countries distributed throughout the similarity space. Instead, the angular distances for Fula are relatively small. Fula, a Senegambian language spoken primarily in West and Central Africa, is concentrated in a specific region of our embedding space.
    Similarly, we can examine the linguistic diversity within each country. Following the previous approach, we selected the top 15 countries with the highest degree and computed the angular distances to each neighboring language. In a given country, if the languages are more similar, their angular positions in the similarity space should be more concentrated. Figure~\ref{fig:unicodelang_profile}b shows that India and Canada exhibit a broader distribution of angular distances, reflecting the presence of a diverse array of languages. In contrast, Russia and Cameroon display a narrower distribution, indicating a more homogeneous set of languages. Notably, all these countries are linguistic hubs, with approximately 30 or more languages spoken in each. These results suggest that our embedding can serve as an indicator of a language's international reach and linguistic diversity, which is not solely reflected by its degree.

    In addition, we investigated the human metabolic network, defined as metabolites connected to the reactions they participate in~\cite{king2016bigg} and the network of food ingredients based on the flavor compounds they share~\cite{ahn2011flavor} (see Section~\ref{sec:datasets}). In both cases, B-Mercator is able to reproduce topological properties of this bipartite network such as the degree distributions and clustering spectra. See Supplementary Note 6 for more details.
    
    Finally, we applied the bipartite greedy routing protocol to the embeddings derived from these real-world networks. For the Unicodelang network, the highest success rate—based on the $A-A$ BGR variant—is observed at an embedding dimension of $D=4$. However, variations in the success probability ($p_s$) across different embedding dimensions are minimal, likely due to the relatively low value of $\beta_b$. In contrast, the highest $p_s$ for both the Metabolic and Flavor networks is achieved at $D=1$. A summary of these findings is provided in Supplementary Tables S2--S5.

    \subsection{Case study on the graph machine learning tasks}
    \label{sec:ml}
    
    Graph Machine Learning (Graph ML) focuses on extracting patterns, making predictions, and uncovering insights from graph-structured data~\cite{hu2020open,xia2021graph}. This data is typically defined as a set of entities (nodes) with complex relationships (links), defining a unipartite graph $\mathcal{G}_n$. Nodes are classified in different categories (labels) and are enriched with a set of features, defining a bipartite network of nodes and features $\mathcal{G}_{n,f}$. Within the Graph ML community, two key tasks are commonly used to evaluate and rank network embedding models: node classification (NC) and link prediction (LP). In turn, network embeddings can be broadly categorized into supervised and unsupervised approaches. Supervised embeddings, such as those learned by graph neural networks (GNNs), use node labels in the training set to inform the learning process for classification tasks. In contrast, unsupervised methods leverage only the network structure and, optionally, node features to generate low-dimensional representations of the data. These maps can then be used for multiple downstream tasks by integrating additional classification models.
    
    \begin{figure*}[t!]
        \centering
        \includegraphics[width=\textwidth]{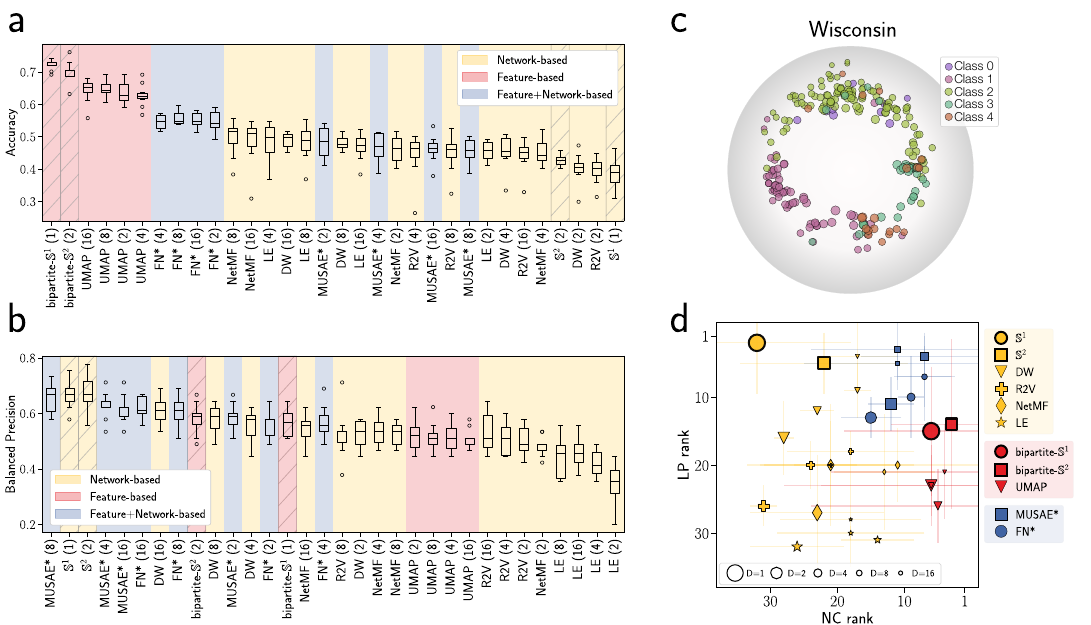}
        \caption{\textbf{Case study on machine learning tasks}. (\textbf{a}) Accuracy of the node classification (NC) task and (\textbf{b}) balanced precision of the distance-based link prediction (LP) task for the Wisconsin dataset. The train/test split is 20/80 for nodes in NC and 90/10 for links in LP, where the test set is balanced by randomly adding an equal number of negative links selected from non-existing links. The results are averaged over 10 different splits. Our methods are highlighted with diagonal hatches. The abbreviations of the algorithms are as follows: DW – DeepWalk, R2V – Role2Vec, LE – Laplacian Eigenmaps, FN – FeatherNode. With a star, we highlighted methods that use both network topology and node features for embedding. The numeric value in brackets indicates the embedding dimension. All other parameters are set to their default values. The methods are sorted by the median accuracy in NC and balanced precision in LP tasks. In addition, the methods are grouped and colored by input data type, i.e., Network-based methods use only network topology, Feature-based methods use only nodes' feature matrix, and Feature$+$Network-based methods merge two things to construct the network embedding. (\textbf{c}) Visualization of the bipartite embedding with B-Mercator in $D=1$ for the Wisconsin dataset. We plot the positions of the nodes and color them based on the metadata. (\textbf{d}) Average rank in link prediction and node classification tasks across seven datasets. We plot the median rank value with the error bars depicting the interquartile range (IQR), i.e., the interval between the first and third quartiles that captures the middle 50\% of the rank distribution. The size of the markers is inversely proportional to the embedding dimension and their shapes correspond to the input data type. Our methods are outlined with a thicker marker border.}
        \label{fig:ml}
    \end{figure*}

    Our embedding method B-Mercator belongs to the class of unsupervised graph embeddings that leverage only the nodes' feature matrices. This is possible thanks to the findings in~\cite{aliakbarisani2023feature}, which show that nodes and their associated features define a bipartite network, $\mathcal{G}_{n,f}$, with strong geometric properties. We used B-Mercator to find an embedding of the nodes-features bipartite network in the common similarity space, which we subsequently used to perform NC and LP tasks in a supervised manner. To highlight the importance of such embeddings, we compared B-Mercator with D-Mercator~\cite{jankowski2023dmercator}, which produces multidimensional hyperbolic maps of unipartite networks $\mathcal{G}_{n}$ without using information from the nodes' features. It has been shown that real complex networks can be accurately embedded in low dimensional hyperbolic spaces~\cite{jankowski2024feature}. Thus, in this work we selected embedding dimensions $D=1$ and $D=2$ to map the node features into the bipartite-$\mathbb{S}^1$ and bipartite-$\mathbb{S}^2$ models using B-Mercator, and to map the unipartite network into the $\mathbb{S}^1$ and $\mathbb{S}^2$ models using D-Mercator.
    
    We also compared our model-driven approach with existing state-of-the-art graph embedding methods in both node classification and link prediction tasks. Prior work shows that efficient encoding can often be achievable in low-dimensional spaces \cite{gu2021principled}. Therefore, we consider embedding dimensions between 2 and 16. For further details on the methods used, see Supplementary Note~7. We selected seven graph datasets commonly used in machine learning research, each with varying levels of correlations between the graph, nodes' features, and nodes' labels (see Supplementary Note~8 for details). It has been shown recently that the performance metrics of Graph ML tasks can vary significantly depending on these correlations~\cite{jankowski2024feature}. For instance, in the node classification tasks, adding features can be detrimental when the correlation between node features and network structure is very low. In contrast, adding features significantly enhances the results when the correlation is high. Among the analyzed networks, Cora and Citeseer exhibit a strong correlation between network structure and node features. In contrast, IMDB, Wisconsin, Texas, and Cornell show relatively low correlation, while Film demonstrates almost no correlation (see Supplementary Table S6 for more details).

    To compute the accuracy of the node classification task, we applied a KNeighborsClassifier from the scikit-learn library~\cite{pedregosa2011scikit} to each network embedding with $K=10$ as the number of nearest neighbors. We split the data into training and testing subsets with a 20/80\% ratio. In the case of B-Mercator and D-Mercator, we computed distances among pairs of nodes using their angular separation on the $D$-sphere. For the rest of the methods, we computed the Euclidean distance between the nodes' positions from the corresponding embedding. In Fig.~\ref{fig:ml}a, we report the performance of the NC task on the Wisconsin dataset. This is a network of web pages from the Computer Science department of the University of Wisconsin. Each web page is enriched with 1613 features and manually classified into one of five categories: student, project, course, staff, and faculty. For this dataset, feature-based methods demonstrate superior performance, with our methods (B-Mercator in $D=1$ and $D=2$) achieving a significant margin of improvement over competitors. This is due to the strong correlation between the nodes' features and labels, as shown in~\cite{jankowski2024feature}. In contrast, network-based methods tend to perform less effectively, as they rely solely on the graph topology, which is weakly correlated with the nodes' labels~\cite{jankowski2024feature}.

    For the distance-based link prediction task, we randomly selected a small fraction \( q = 0.1 \) of the existing links from the network structure as positive links. Similarly, an equal number of non-existing edges was selected as negative links to form a balanced test set. The remaining positive links were considered as the training network, which was then embedded using various network-based methods. In the case of B-Mercator, the bipartite network between nodes and features was embedded, and it was not affected by splitting the links in the network structure into training and test sets. For B-Mercator and D-Mercator, we used the inverse of the hyperbolic distance between node pairs as the similarity measure, while for the other methods, we used the inverse of the Euclidean distance between node positions. As a result, links in the test set with the smallest distance were ranked highest when sorted in ascending order of their similarity scores.
    
    In Fig.~\ref{fig:ml}b, we show the precision of the different methods. In this case, methods using only the network topology, or a combination of network structure and node features, deliver the highest precisions. For this task, our embeddings with D-Mercator ($D=1$ and $D=2$), using only the network topology, are competitive and achieve a performance similar to MUSAE in dimension $D=8$, demonstrating the versatility of low-dimensional embeddings derived from our approach. Moreover, among feature-based methods, B-Mercator in $D=1$ and $D=2$ outperforms all other approaches. This highlights the adaptability of our methods to different tasks depending on the underlying data representation and task requirements. Detailed results for additional datasets can be found in Supplementary Figures S25–S50.
    
    Figure~\ref{fig:ml}d provides a summary of the results across multiple datasets. For each dataset, the rank of each method is calculated for both NC and LP tasks, and the median rank is plotted along with the interquartile range (IQR). B-Mercator in low dimensions ($D=1$ and $D=2$) consistently ranks among the top approaches for the NC task and outperforms all feature-based methods for the LP task, for which D-Mercator is the best method in dimension $D=1$. These results demonstrate the reliability and adaptability of our methods across datasets. The marker size in the plot, which inversely represents the embedding dimension, further underscores the efficiency of our low-dimensional embeddings compared to high-dimensional alternatives. In Supplementary Figures S32 and S42, we provide a more detailed view of the rankings for the NC and LP tasks.

    \section{Discussion}
    
    The ability to embed real-world systems into a geometric space is a pivotal step toward understanding their intrinsic structure, function, and underlying organization. While numerous network embedding techniques have been proposed, most are not derived from a model-based perspective, limiting their interpretability and capacity to reconstruct empirical data. Model-based approaches offer a principled way of capturing the generative mechanisms that shape network topologies, thereby enabling researchers to interpret embeddings in a manner closely aligned with the data's underlying structure.
    
    Several model-based embedding methods have already proved effective for unipartite networks, yet comparable solutions for bipartite networks have remained underexplored. In this work, we addressed this gap by introducing B-Mercator, a geometric model-based embedding algorithm specifically designed for bipartite networks. By mapping bipartite structures into hyperbolic space, B-Mercator offers a powerful and interpretable way to capture community structure, hierarchical organization, and topological relationships.
    
    To demonstrate the versatility of B-Mercator, we applied it to embed real-world bipartite systems. The analysis of the language network (countries--spoken languages), the metabolic network (metabolites--reactions), and the Flavor network (ingredients--chemical compounds) show that the embeddings not only correlate well with metadata but also retain the essential characteristics of each dataset. Furthermore, we evaluated B-Mercator on several datasets on the node classification and link prediction tasks. It consistently outperformed all unsupervised methods for node classification and emerged as the best performer among feature-based embeddings for link prediction.
    
    We stress an important advantage of our model-based embeddings. The hyperbolic maps can be used to generate synthetic networks with node features that closely resemble their real-world counterparts (see Supplementary Figures 51-53). By doing so, we safeguard any sensitive information derived from real complex networks, such as personal connections, transactional data, or proprietary interactions. Thus, we enable the secure sharing of structural data without compromising the integrity of the original network or revealing sensitive information.
   
    These findings underscore the value of geometric model-based embeddings for both theoretical analyses and practical applications, ranging from community detection and studying network hierarchies to advanced machine learning tasks. B-Mercator's robust performance highlights its capacity to reveal meaningful insights into bipartite systems --a domain often overlooked in the current embedding literature-- while providing substantially more accurate analyses than those based on corresponding one-mode projections. Overall, B-Mercator represents a significant advancement in bipartite network analysis, paving the way for more accurate, interpretable, and generative representations of complex real-world systems.
    
    \vfill
    
    \section{Methods}
    
    \subsection{Bipartite-$\mathbb{S}^D/\mathbb{H}^{D+1}$ model}
    \label{bipartiteS1H2}
    In the bipartite-$\mathbb{S}^D/\mathbb{H}^{D+1}$ model -- an extension of the bipartite-$\mathbb{S}^1/\mathbb{H}^{2}$~\cite{aliakbarisani2023feature} -- we assign to each node (of type-A or type-B) a hidden degree ($\kappa_A$ or $\kappa_B$) and the position in the $D$-dimensional similarity space chosen uniformly at random, and represented as a point on a $D$-dimensional sphere. Each node of type-A and -B is assigned a vector $\mathbf{x}_{i} \in \mathbb{R}^{D+1}$ with $||\mathbf{x}_{i}||=R$. For instance, when $D=1$ the similarity space is represented as a circle, whereas for $D=2$ it is a sphere (see Figure~\ref{fig:schematic}). 
    
    The connection probability between type-A node $u$ and type-B node $v$ takes the form of gravity law:
    \begin{align}
        \label{eq:prob}
        {p}_{uv}=\frac{1}{1+{\chi}_{uv}^{\beta_b}},\,\,\,\mathrm{with} \,\,\,{\chi }_{uv}=\frac{R{{\Delta }}{\theta }_{uv}}{{\left(\mu_b {\kappa}_{u}{\kappa }_{v}\right)}^{1/D}}.
    \end{align}
    The number of type-A (type-B) nodes is $N_A$($N_B$), for convenience and without loss of generality, we set the density of type-A nodes in the $D$-sphere to one so that 
    \begin{align}
        R={\left[\frac{N_A}{2{\pi }^{\frac{D+1}{2}}}{{\Gamma }}\left(\frac{D+1}{2}\right)\right]}^{\frac{1}{D}}.
    \end{align}
    The separation $\Delta \theta_{uv} = \arccos \left(\frac{\mathbf{x}_{u}.\mathbf{x}_{v}}{R^2}\right)$
    represents the angular distance between nodes $u$ and $v$ in the $D$-dimensional similarity space. The parameter $\beta_b$ (with $\beta_b>D$) controls the coupling between the resulting topology and the underlying metric space. Lastly, the parameter $\mu_b$ controls the average degree of type-A nodes and is defined as
    \begin{align}
        \label{eq:mu}
        \mu_b=\frac{\beta_b {{\Gamma }}\left(\frac{D}{2}\right)\sin \frac{D\pi }{\beta_b }}{2{\pi }^{1+\frac{D}{2}} \langle k_A \rangle },
    \end{align}
    whereas the average degree of type-B nodes is set by $\langle k_B \rangle = \frac{N_A}{N_B} \langle k_A \rangle$. By choosing distributions for the hidden degrees $\kappa_A$ and $\kappa_B$ and inverse temperature $\beta_b$, we can generate bipartite networks with any desired degree distributions and varying levels of geometric properties.
    
    The bipartite-$\mathbb{S}^D$ model can be represented in purely geometric terms as the bipartite-$\mathbb{H}^{D+1}$ model. This is achieved by mapping the hidden degrees of each type-A and -B nodes to radial coordinates while preserving their positions on the $D$-sphere. Specifically, the transformation for type-A nodes has the form (similarly for type-B nodes)
    \begin{align}
        {r}_{u}=\hat{R}-\frac{2}{D}\ln \frac{{\kappa }_{u}}{{\kappa }_{u, 0}},\,\,\,{{\rm{with}}}\,\,\,\hat{R}=2\ln \left(\frac{2R}{{(\mu_b {\kappa }_{u, 0}{\kappa }_{v, 0})}^{1/D}}\right).
    \end{align}
    where ${\kappa }_{u, 0}$(${\kappa }_{v, 0}$) is the smallest hidden degree for type-A (type-B) nodes. We can rewrite Eq.~\eqref{eq:prob} as
    \begin{align}
        {p}_{uv}=\frac{1}{1+{e}^{\frac{\beta_b }{2}({x}_{uv}-\hat{R})}},\,\,\,{{{{{{{\rm{with}}}}}}}}\,\,\,{x}_{uv}={r}_{u}+{r}_{v}+2\ln \frac{{{\Delta }}{\theta }_{uv}}{2}.
    \end{align}
    With this transformation, the space represented by the radial position of each node, along with its angular position on the sphere, becomes the native representation of the hyperbolic space of dimension $D+1$. Consequently, the connection probability between nodes $u$ and $v$ becomes a function of $x_{uv}$, which is a good approximation of the hyperbolic distance between them~\cite{serrano_boguna_2022}.

    \subsection{B-Mercator in detail}
    \label{BMercator}
    We adopt the code of $D$-Mercator~\cite{jankowski2023dmercator} for embedding bipartite networks. Here, we provide an overview of the differences between the embeddings for the unipartite and bipartite networks. See also Supplementary Note~12 for more technical details.
    
    \paragraph{Inferring the hidden degrees and parameter $\beta_b$.} 
    The inference of hidden degrees for type-A and -B nodes, and the inverse temperature $\beta_b$ is implemented as an iterative process. We begin with the initial guess for the parameter $\beta_b \in (D, 2D)$ where $D$ is the embedding dimension, and initialize the hidden degrees as the observed degrees in the original network for type-A ($k_A = \{k_{A,i}\}_{i=1}^{N_A}$) and type-B  ($k_B = \{k_{B,j}\}_{j=1}^{N_B}$) nodes. The aim of the estimation is to modify the hidden degrees in order to ensure that the expected degree of each node within the model aligns with the degree observed in the original network. After the hidden degrees for both nodes A and B are computed, the synthetic graph from the  bipartite-$\mathbb{S}^D/\mathbb{H}^{D+1}$ is constructed and the bipartite clustering coefficient is calculated. If the computed bipartite clustering coefficient deviates from that of the original network, $\bar{c}_b$, the value of $\beta_b$ is adjusted. Then, the process is repeated using the current estimation of hidden degrees until a predetermined precision is reached.
    
    \paragraph{Bipartite-$\mathbb{S}^D$ model corrected Laplacian Eigenmaps.}
    Since the biadjacency matrix of a bipartite graph $\mathbf{A}$ is not symmetric, to apply the Laplacian Eigenmaps~\cite{belkin2003laplacian}, we transform it to the adjacency matrix as
    \begin{align}
        \mathbf{B} = \begin{bmatrix}
            0 & \mathbf{A} \\
            \mathbf{A}^T & 0 
        \end{bmatrix}
    \end{align}
    Similarly to \cite{jankowski2023dmercator}, the expected angular distance between nodes $u$ (of type-A) and $v$ (of type-B) in the bipartite-$\mathbb{S}^D$ model, conditioned to the fact that they are connected, can be computed as
    \begin{align}
        \langle \Delta \theta_{uv}\rangle &= \int\limits_{0}^{\pi} \Delta\theta_{uv} \rho(\Delta\theta_{uv}|a_{uv} =1) \, d\Delta\theta_{uv}
    \end{align}
    Additionally, for $D=1$, we keep the ordering inferred by LE and distribute type-A and type-B nodes evenly on the circle.
    
    \paragraph{Likelihood maximization}
    The nodes’ coordinates in the similarity space inferred using LE are adjusted by Maximum Likelihood Estimation (MLE) to optimize the probability that the bipartite-$\mathbb{S}^D$ model generates the observed network. We define an order of nodes sorted by their degree for A and B type nodes separately. Fixing the positions of a subset of B type nodes, we find new optimal coordinates for a subset of type-A nodes. First, we compute the mean coordinates of type-A node $u$'s neighbors. 
    \begin{align}
        \mathbf{x}_{u} = \sum\limits_{v} \frac{1}{\kappa_v^2} \mathbf{x}_v
    \end{align}
    where the sum goes over all neighbors of node $u$, i.e., type-B nodes. Later, the new positions around $\mathbf{x}_u$ are proposed using a multivariate normal distribution. Finally, we select the most likely candidate position based on the local log-likelihood
    \begin{align}
        \ln \mathcal{L}_u = \sum\limits_{v = 0}^{N_B} a_{uv} \ln p_{uv} + (1 - a_{uv}) \ln (1 - p_{uv})
    \end{align}
    After iterating over a subset of type-A nodes, we apply a similar approach to type-B nodes, and repeat the process until all node positions are adjusted.
    
    \paragraph{Final adjustment of hidden degrees}
    Lastly, we adjust the hidden degrees to compensate deviations from $\bar{k}_A(\kappa_u) = \kappa_u$ and $\bar{k}_B(\kappa_v) = \kappa_v$, which might have been introduced during the estimation of the nodes' coordinates in the similarity space.
    
    \subsection{Real bipartite datasets}
    \label{sec:datasets}
    The Unicodelang dataset was downloaded from the Unicode CLDR Project GitHub repository~\cite{unicode_cldr}. The dataset~\cite{githubCldrcommonsupplementalsupplementalDataxmlMain} contains information about the languages spoken in a given territory. We preprocessed this dataset and matched the country codes to their geographical regions~\cite{githubGitHubLukesISO3166CountrieswithRegionalCodes}. In total, the bipartite graph contains 246 countries, 717 languages, and 1487 edges. 
    
    The metabolic network was extracted from the BiGG webpage~\cite{king2016bigg}. We focus on the RECON1 model, which corresponds to \textit{Homo sapiens}. The data was preprocessed and cleaned, resulting in 1497 metabolites and 2212 chemical reactions.
    
    The flavor network is a network of food ingredients based on the flavor compounds they share~\cite{ahn2011flavor}. After preprocessing, the total number of ingredients is 602, and the number of compounds is 1138. The bipartite graph contains 15,382 edges. Additional network properties and inferred values of $\beta_b$ for different embedding dimensions are shown in Supplementary Table 1.
    
    \section*{Acknowledgments}
    We acknowledge support from: Grant TED2021-129791B-I00 funded by MCIN/AEI/10.13039/501100011033 and the ``European Union NextGenerationEU/PRTR''; Grant PID2022-137505NB-C22 funded by MCIN/AEI/10.13039/501100011033 and by ``ERDF/EU''. R.~J. acknowledges support from the fellowship FI-SDUR funded by Generalitat de Catalunya. M. B. acknowledges the ICREA Academia award, funded by the Generalitat de Catalunya.
    
    \section*{Author Contributions}
    
M. B., M. A. S., R. J., and R. A. designed research. R. J., and R. A. performed research. M. B., M. A. S., R. J. wrote the paper. All authors discussed results, reviewed the manuscript, and approved the final version.
    
    \section*{Competing interests}
    All authors declare no competing interests.
    
    \section*{Data availability}
    The network datasets used in this study are available from the sources referenced in the manuscript and the Supplementary Information.
    
    \section*{Code availability}
    The open-source code for B-Mercator, along with the code to reproduce the figures, is available on GitHub~\cite{bmercator_github}.

    \section*{References}

    \newpage
    

\section*{Supplementary material}

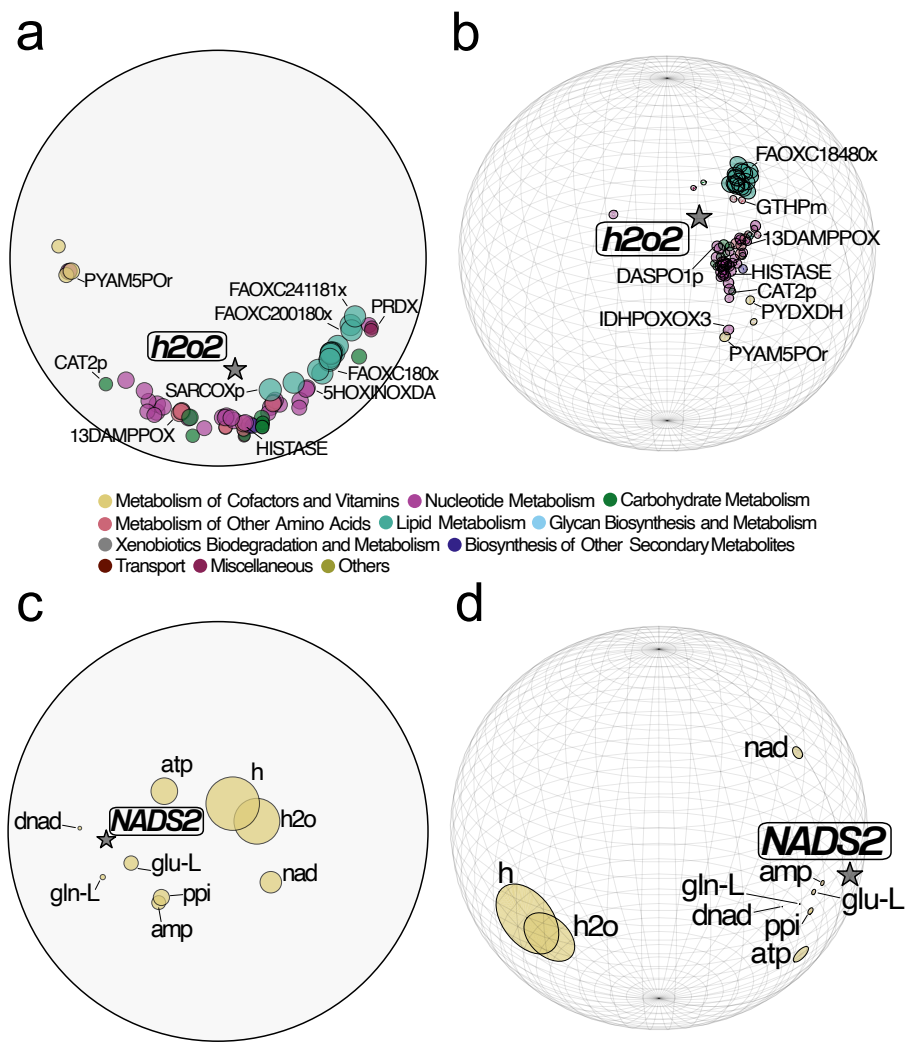

FIG. S16: **Visualization of the bipartite hyperbolic embeddings of the Metabolic dataset per metabolite or reaction in** $D = 1$ **and** $D = 2$. Panels (**a, b**) show positions of all reactions connected to the $H_2O_2$ metabolite, whereas panels (**c, d**) all metabolites having a link to a NADS2 reaction (Nicotinate-mononucleotide adenylyltransferase). The size of the nodes is proportional to the nodes' degree. The color in panels (**a, b**) corresponds to the reaction category. A star marker indicates the position of a given metabolite or reaction.

Here, we focus on the human metabolic network, defined as metabolites connected to the reactions they participate in [1]. B-Mercator is able to reproduce topological properties of the metabolic network, such as degree distributions and clustering spectra (see Figure S18). Figures S16a,b show all reactions in which the metabolite $H_2O_2$ is present from embeddings for $D = 1$ and $D = 2$. We can distinguish two main reaction clusters corresponding to Nucleotide Metabolism and Lipid Metabolism reaction types. The $H_2O_2$ metabolite is located between these two communities. In Figure S17, we plot the angular distribution for each reaction type for the embedding in $D = 1$. In Figures S16c,d, we plot all metabolites participating in the NADS2 reaction. Nicotinate-mononucleotide adenylyltransferase is a key enzyme that helps produce NAD, a molecule essential for generating energy and supporting various vital processes in cells. One can observe that the metabolites H and $H_2O$ are located close to the center of the hyperbolic disk. These metabolites are hubs in the bipartite network and participate in many chemical reactions.

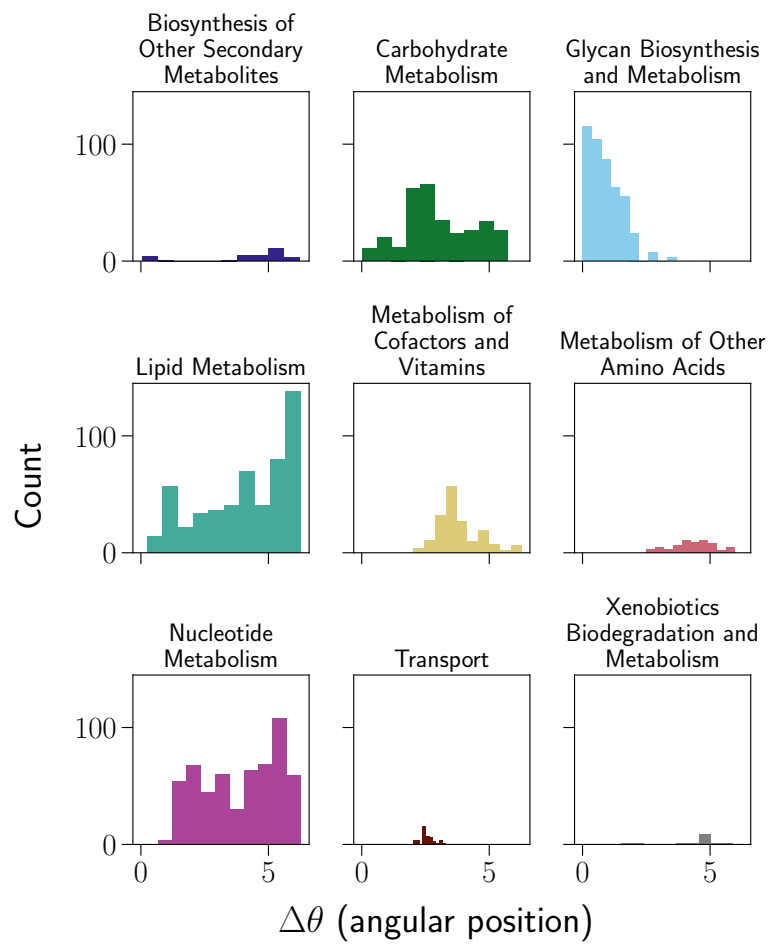

FIG. S17: The angular distribution of reactions (type B nodes) grouped by reaction type in the metabolic dataset. We plot the reaction types consisting more than 15 nodes.

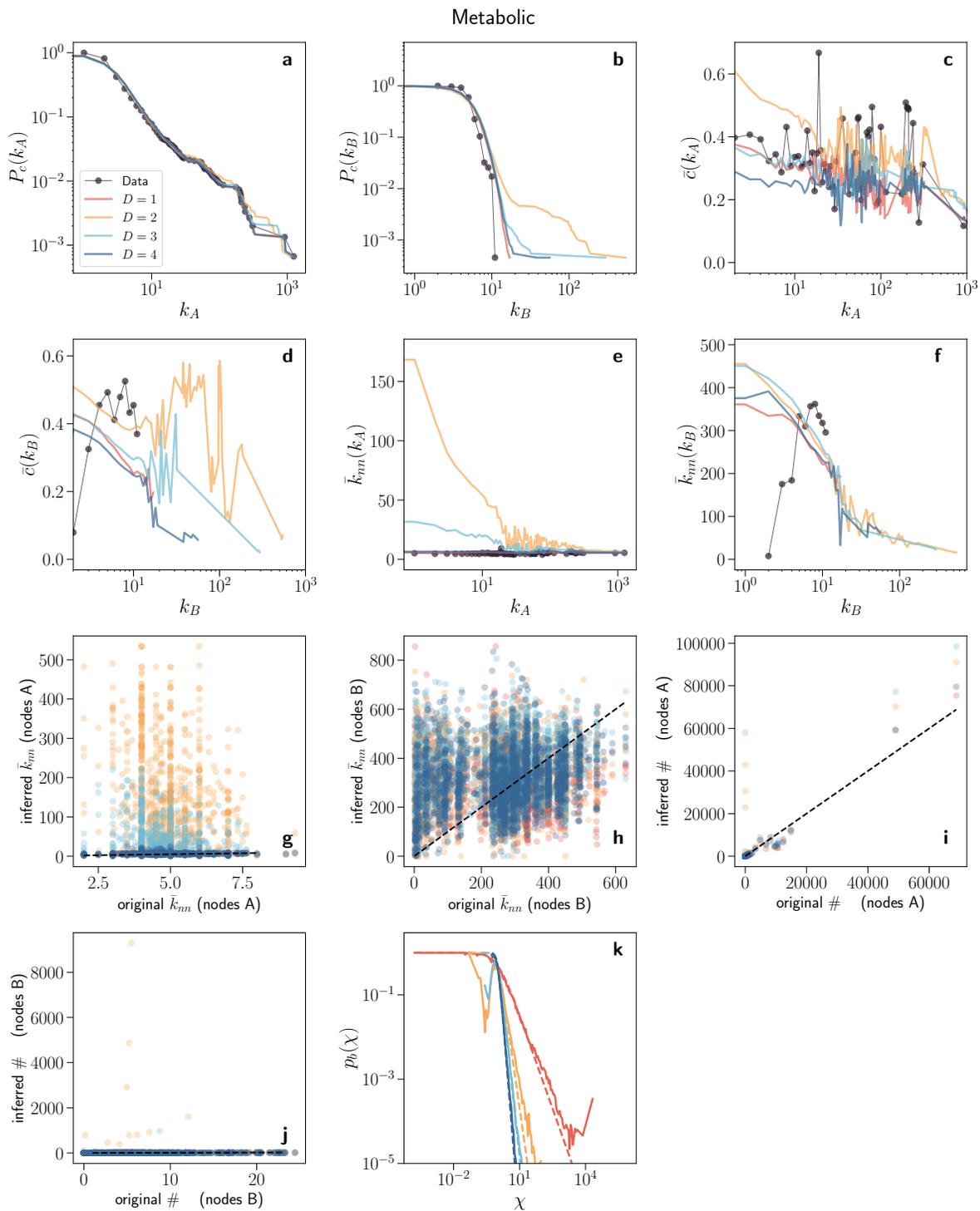

FIG. S18: Topological validation of the Metabolic dataset in which type A nodes are metabolites and type B nodes are reactions. See caption in Fig. S6 for more details.

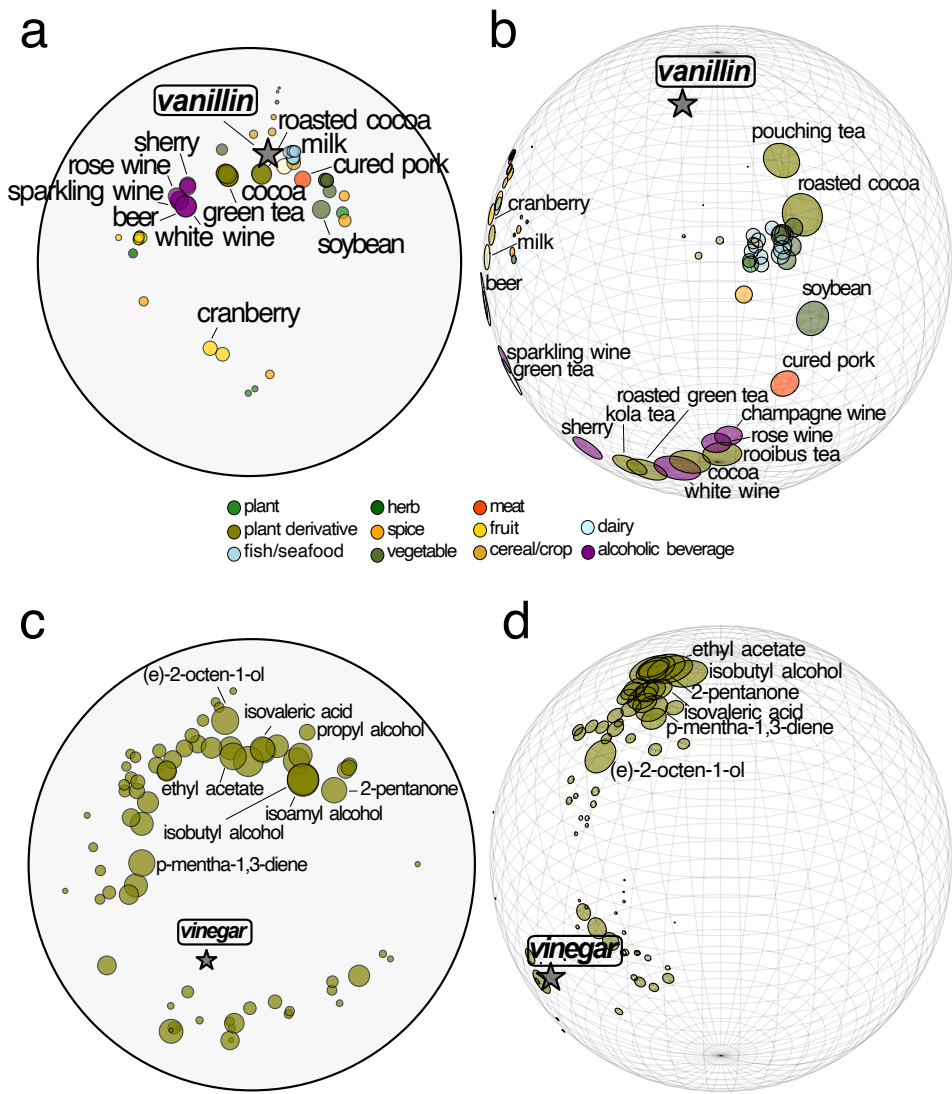

FIG. S19: **Visualization of the bipartite hyperbolic embeddings of the Flavor dataset per ingredient or chemical compound in** $D = 1$ **and** $D = 2$. Panels (**a, b**) show positions of all ingredients connected to the vanillin compound whereas panels (**c, d**) all compounds having a link to a vinegar. The size of the nodes is proportional to the nodes' degree. The color in panels (**a, b**) corresponds to the ingredient category. A star marker indicates the position of a given compound or ingredient.

As an another example, here, we focus on the network of food ingredients based on the flavor compounds they share [2]. In [2], the Flavor network has been analyzed by projecting an ingredient-compound bipartite network into the ingredient space in which nodes are ingredients, linked if they share at least one flavor compound. However, our method works directly on the bipartite network without the need to project it into the unipartite network. This is an key point of our approach since one-mode projections can distort important information of the original bipartite network [3, 4].

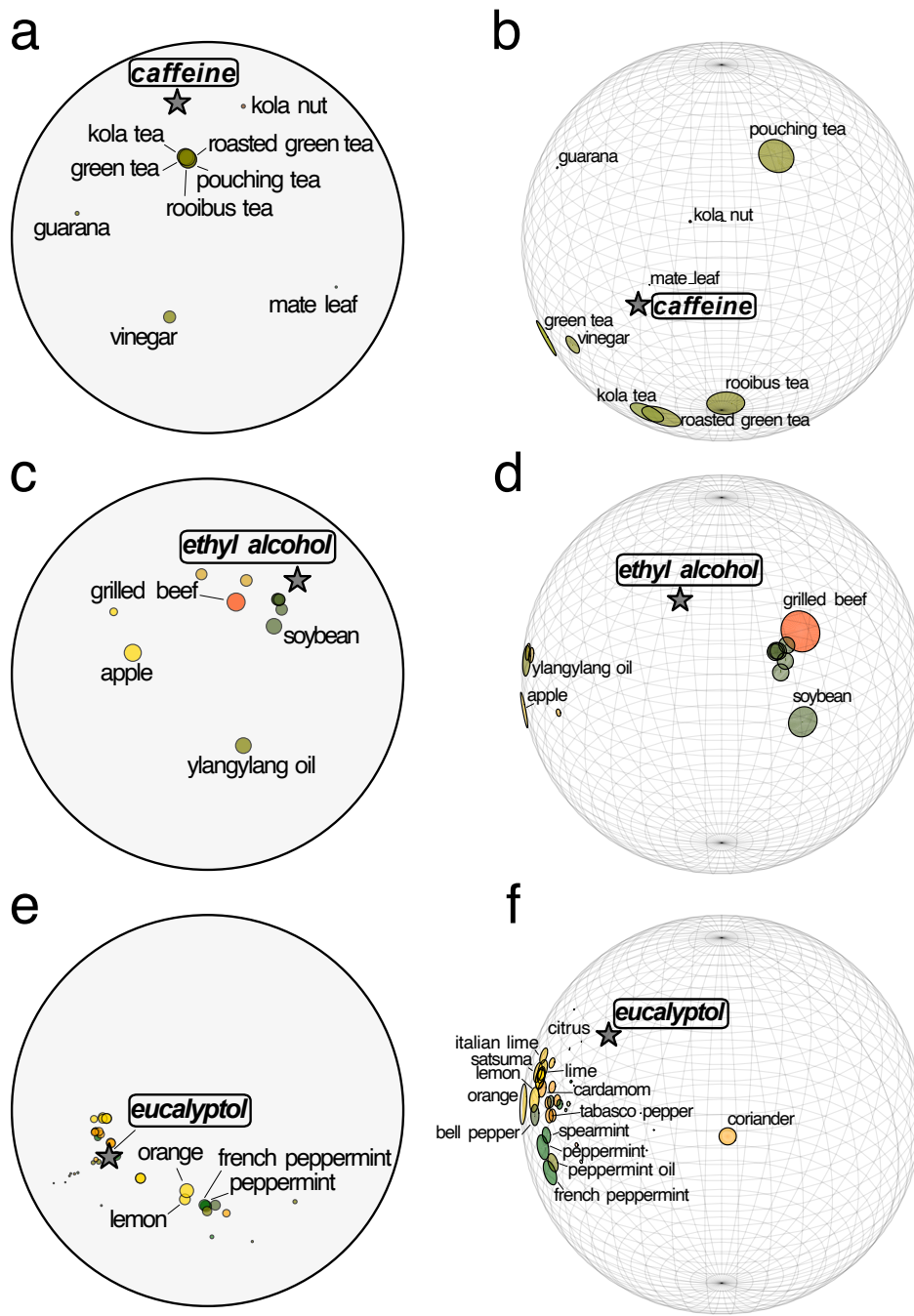

FIG. S20: **Visualization of the $\mathbb{S}^1$ and $\mathbb{S}^2$ embeddings of the Flavour dataset per chemical compound**. Panels show the positions of all ingredients connected to a given chemical compound. The size of the nodes is proportional to the nodes' degree. The color corresponds to the ingredient category. A star marker indicates the position of a given compound.

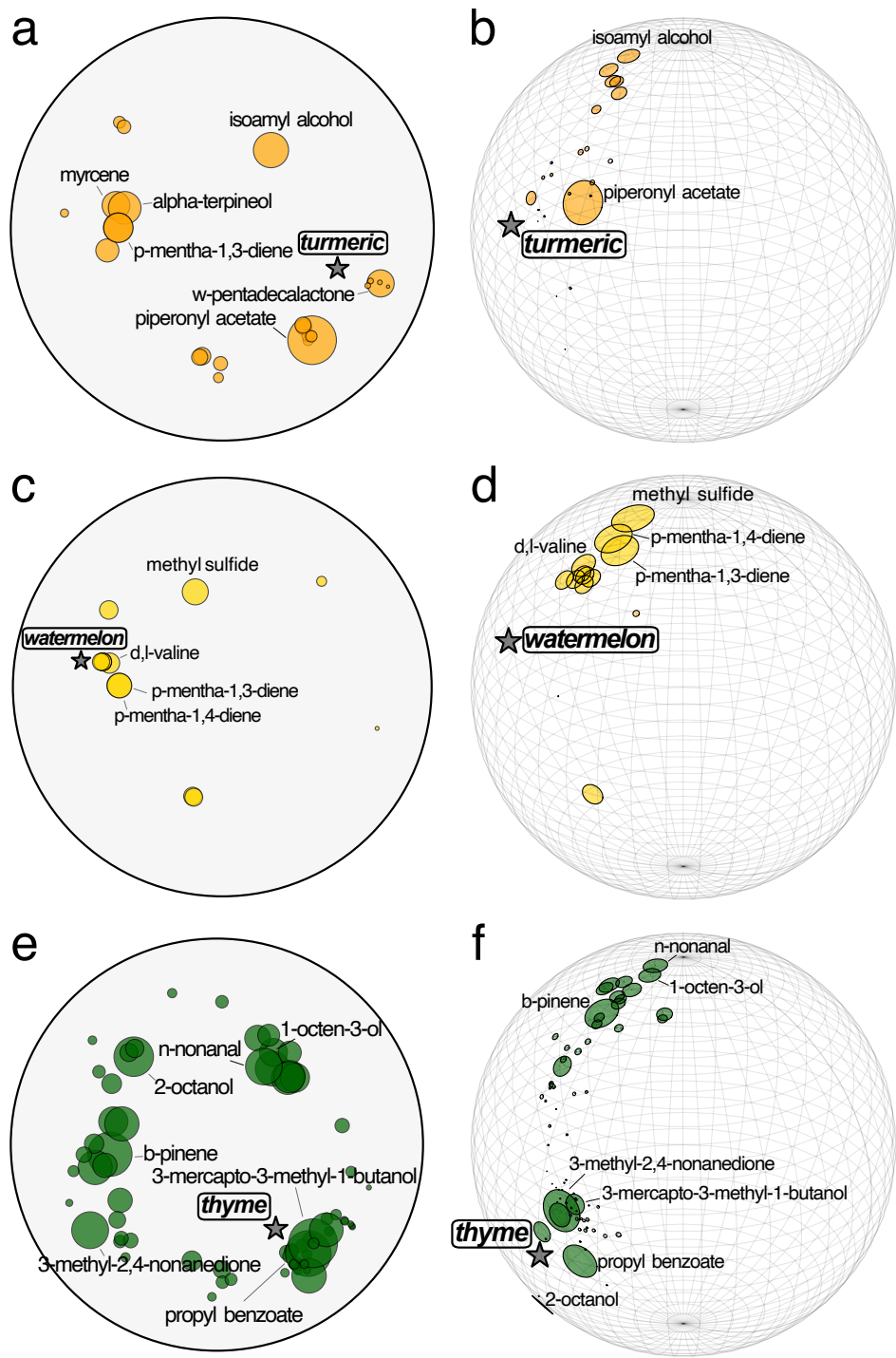

FIG. S21: **Visualization of the $\mathbb{S}^1$ and $\mathbb{S}^2$ embeddings of the Flavour dataset per ingredient**. Panels shows the positions of all chemical compounds connected to a given ingredient. The size of the nodes is proportional to the nodes' degree. A star marker indicates the position of a given ingredient.

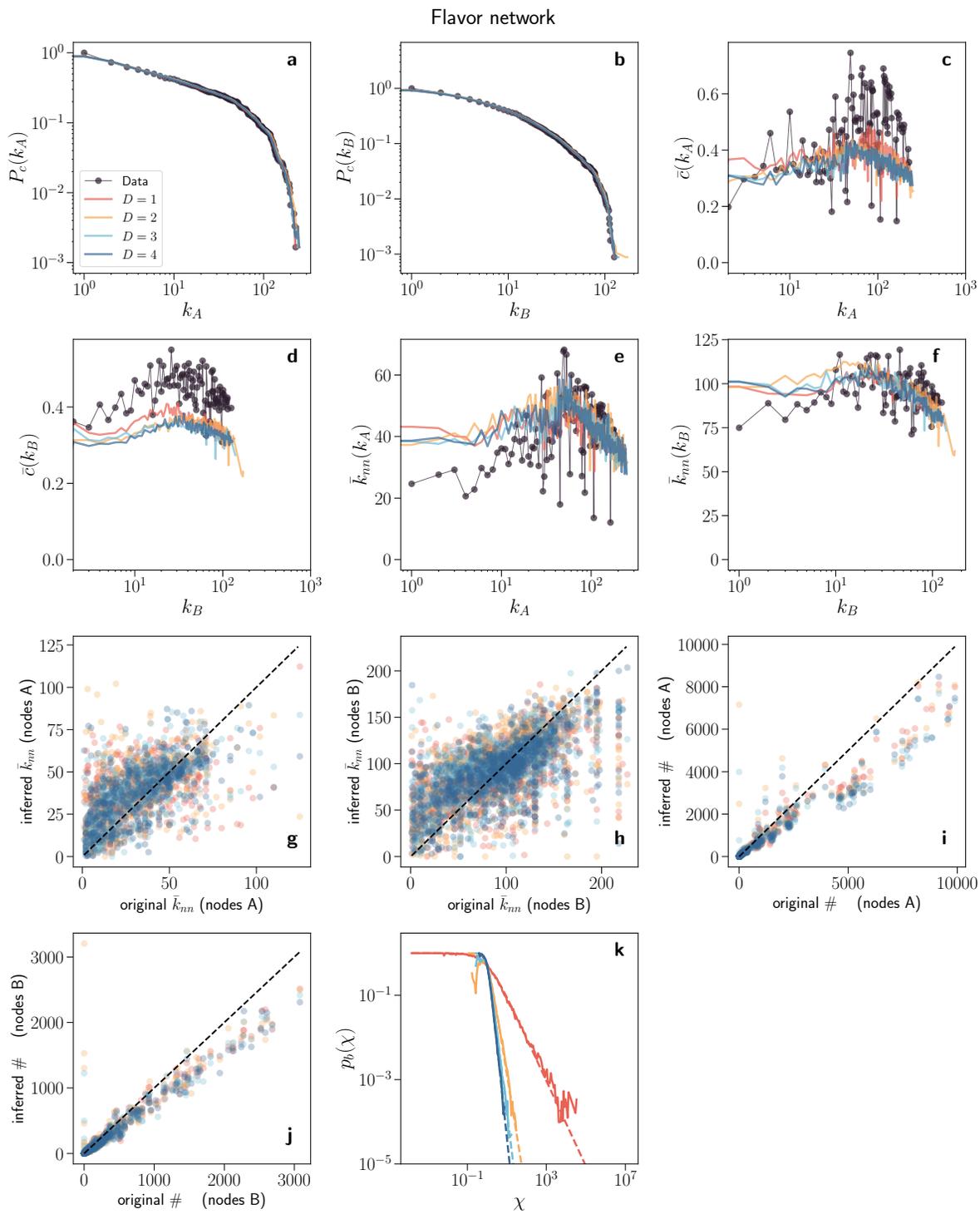

FIG. S22: Topological validation of the Flavor dataset in which type A nodes are ingredients and type B nodes are compounds. See caption in Fig. S6 for more details.

## 7. UNSUPERVISED GRAPH EMBEDDINGS

Here we provide a short summary of each machine learning method.

- DeepWalk [5] uses random walks to approximate the pointwise mutual information matrix obtained by pooling normalized adjacency matrix powers. This matrix is decomposed by an approximate factorization technique.

- Role2Vec [6] uses random walks to approximate the pointwise mutual information matrix obtained by multiplying the pooled adjacency power matrix with a structural feature matrix (in this case Weisfeiler-Lehman features). This way one gets structural node embeddings.

- NetMF [7] uses sparse truncated SVD to learn embeddings for the pooled powers of the pointwise mutual information matrix computed from powers of the normalized adjacency matrix.

- LaplacianEigenmaps [8] extracts the eigenvectors corresponding to the largest eigenvalues of the graph Laplacian. These vectors are used as the node embedding.

- FeatherNode [9] uses characteristic functions of node features with random walk weights to describe node neighborhoods.

- MUSAE [10] performs attributed random walks to approximate the pooled adjacency matrix power node feature matrix product. The matrix is decomposed implicitly by a Skip-Gram style optimization problem.

- UMAP [11] is a dimension reduction technique that takes a node feature matrix and maps it into a low-dimensional Euclidean space.

## 8. MACHINE LEARNING DATASETS

- Film [12]. Actor co-occurrence network. This dataset is the actor-only induced subgraph of the film-directoractor-writer network. Each nodes correspond to an actor, and the edge between two nodes denotes co-occurrence on the same Wikipedia page. Node features correspond to some keywords in the Wikipedia pages. The nodes are classfied into five categories in term of words of actor's Wikipedia.

- IMDB [13]: The Movie-Actor-Movie relation dataset. Movies are categorized into three classes (Action, Comedy, Drama).

- Citeseer [14]: The citation network of Machine Learning papers where each publication is described by a 0 or 1 valued word vector indicating the absence or the presence of the corresponding word from the dictionary. The dictionary consists of 1433 unique words. The publications are classified into six classes.

- Cora [15]: Similar to Citeseer, however the publications are split into seven classes: Case Based, Genetic Algorithms, Neural Networks, Probabilistic Methods, Reinforcement Learning, Rule Learning, Theory.

- Cornell, Wisconsin, Texas [16]: Web graphs crawled from three Computer Science departments in 1998, with each page manually classified into one of seven categories: course, department, faculty, project, staff, student, or other.

| Dataset | $N$ | $N_l$ | $\langle k \rangle$ | $\bar{c}$ | $\beta$ | $N_f$ | $\langle k_n \rangle$ | $\langle k_f \rangle$ | $\bar{c}_{b,n}$ | $\bar{c}_{b,f}$ | $\beta_b$ | corr$(\mathcal{G}, F)$ |
|---|---|---|---|---|---|---|---|---|---|---|---|---|
| Film | 7600 | 5 | 7.02 | 0.10 | 1.0370 | 932 | 5.39 | 43.97 | 0.513 | 0.394 | 1.5207 | 0.039 |
| IMDB | 3228 | 3 | 19.46 | 0.55 | 2.3276 | 2000 | 76.96 | 124.21 | 0.077 | 0.071 | 1.0170 | 0.172 |
| Cora | 2485 | 7 | 4.08 | 0.28 | 1.5686 | 1428 | 18.3 | 31.85 | 0.134 | 0.1 | 1.0094 | 0.650 |
| Citeseer | 2110 | 6 | 3.48 | 0.23 | 1.4672 | 3604 | 32.07 | 18.77 | 0.122 | 0.13 | 1.0108 | 0.763 |
| Wisconsin | 251 | 7 | 3.59 | 0.28 | 1.0074 | 1613 | 95.85 | 14.91 | 0.582 | 0.434 | 1.0419 | 0.201 |
| Texas | 183 | 7 | 3.05 | 0.32 | 1.0071 | 1500 | 83.42 | 10.18 | 0.57 | 0.413 | 1.0112 | 0.116 |
| Cornell | 183 | 7 | 3.03 | 0.29 | 1.0061 | 1582 | 94.21 | 10.90 | 0.56 | 0.399 | 1.0293 | 0.169 |

TABLE S3: Properties of real networks. The $N$ represents number of nodes in the unipartite network, $N_l$ the number of node labels, $\langle k \rangle$ the average degree, $\bar{c}$ the average clustering coefficient and $\beta$ the inferred inverse temperature for $D = 1$. Meanwhile, $N_f$ corresponds to the number of features, $\langle k_n \rangle$ the average number of nodes per feature, $\langle k_f \rangle$ the average number of features per node. The $\bar{c}_{b,n}$ ($\bar{c}_{b,f}$) is the bipartite clustering for nodes (features). The $\beta_b$ is the inferred inverse temperature for the bipartite network in $D = 1$. Lastly, corr$(\mathcal{G}, F)$ is a measure of correlation between network structure and nodes' features defined in [17]. The higher the obtained value the more correlated are features with the network topology.

## 9. NODE CLASSIFICATION

We perform a node classification task on popular machine learning datasets.

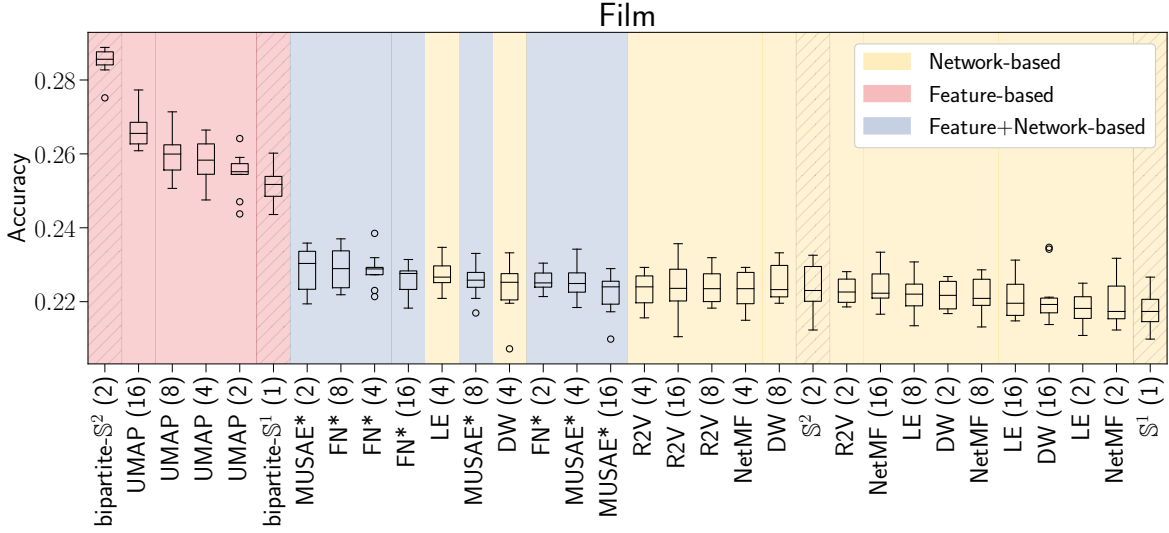

FIG. S23: Accuracy of the node classification task for Film dataset. For each algorithm, we use a KNeighborsClassifier with $K = 10$. The train/test split is 20/80, and the results are averaged over 10 different splits. Our methods are highlighted with diagonal hatches. The abbreviations of the algorithms are as follows: DW – DeepWalk, R2V – Role2Vec, LE – Laplacian Eigenmaps, FN – FeatherNode. The numeric value in brackets indicates the embedding dimension. All other parameters are set to their default values.

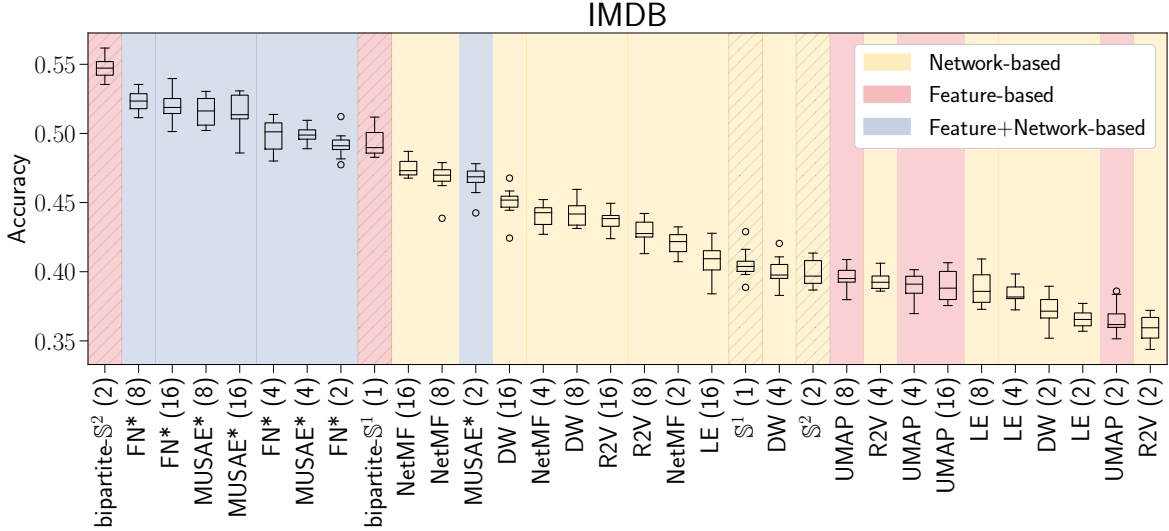

FIG. S24: Accuracy of the node classification task for IMDB dataset. See caption in Fig. S23 for more details.

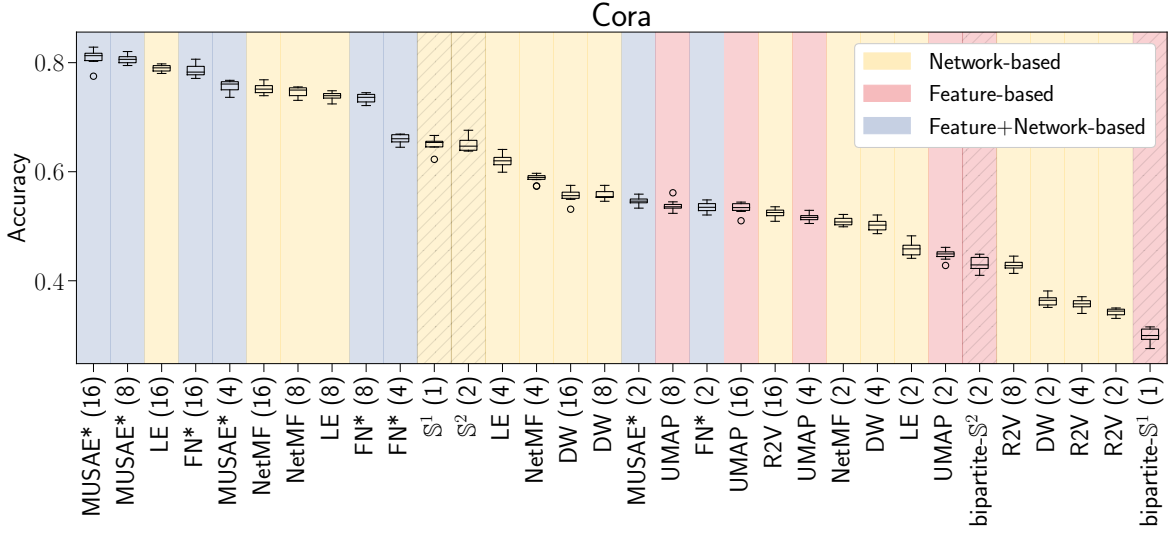

FIG. S25: Accuracy of the node classification task for Cora dataset. See caption in Fig. S23 for more details.

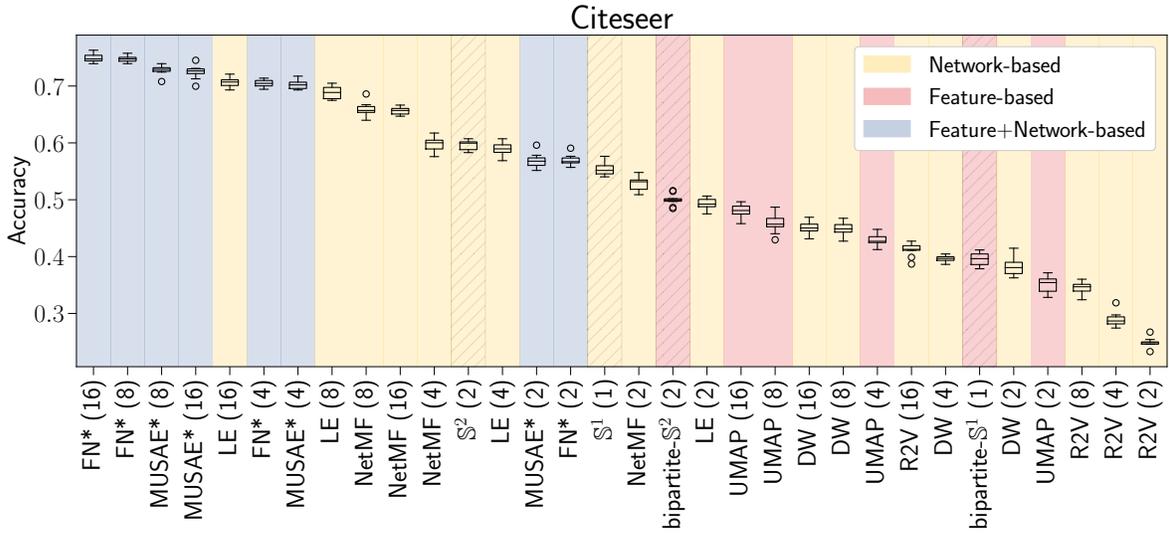

FIG. S26: Accuracy of the node classification task for Citeseer dataset. See caption in Fig. S23 for more details.

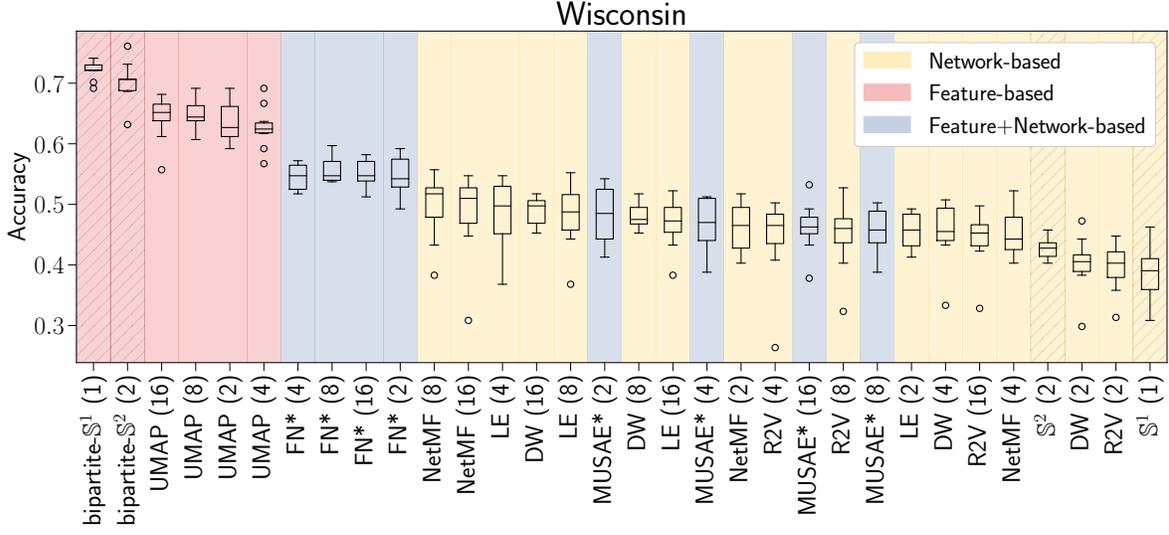

FIG. S27: Accuracy of the node classification task for Wisconsin dataset. See caption in Fig. S27 for more details.

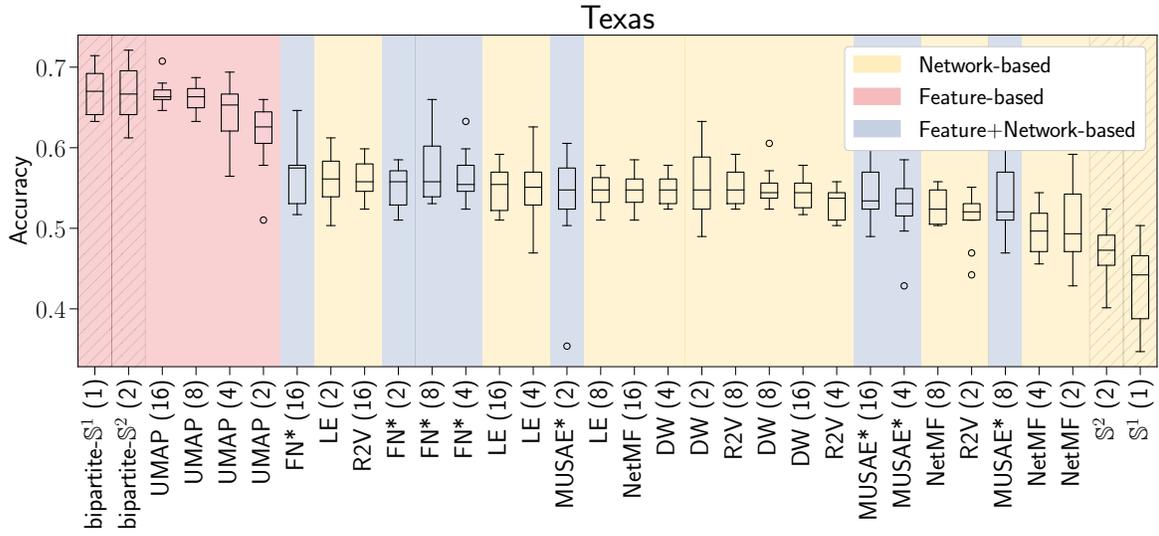

FIG. S28: Accuracy of the node classification task for Texas dataset. See caption in Fig. S23 for more details.

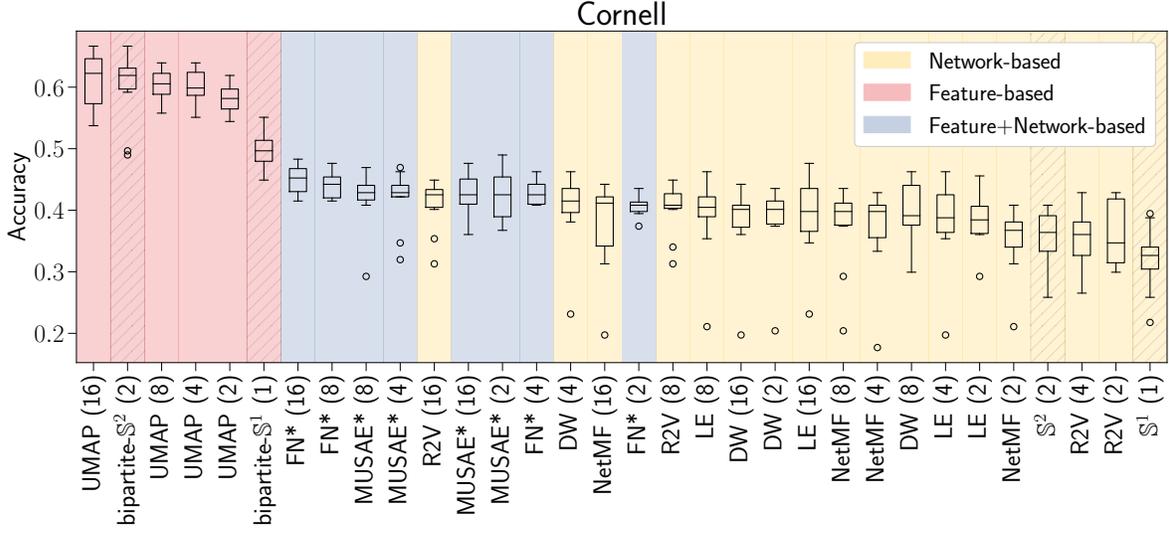

FIG. S29: Accuracy of the node classification task for Cornell dataset. See caption in Fig. S23 for more details.

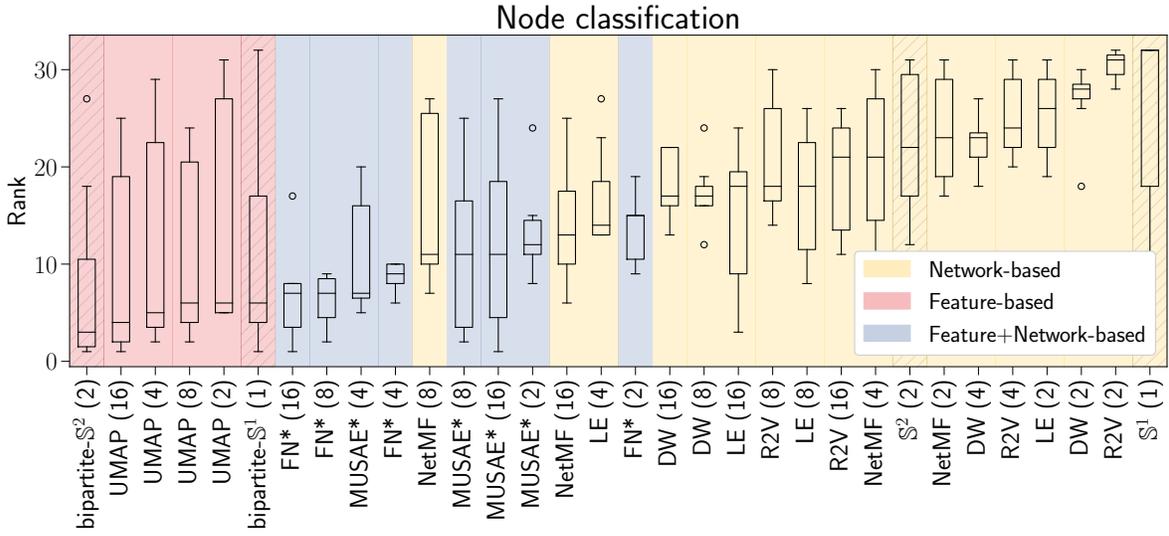

FIG. S30: The rank of network embedding methods across all datasets for the node classification task.

## 10. DISTANCE-BASED LINK PREDICTION

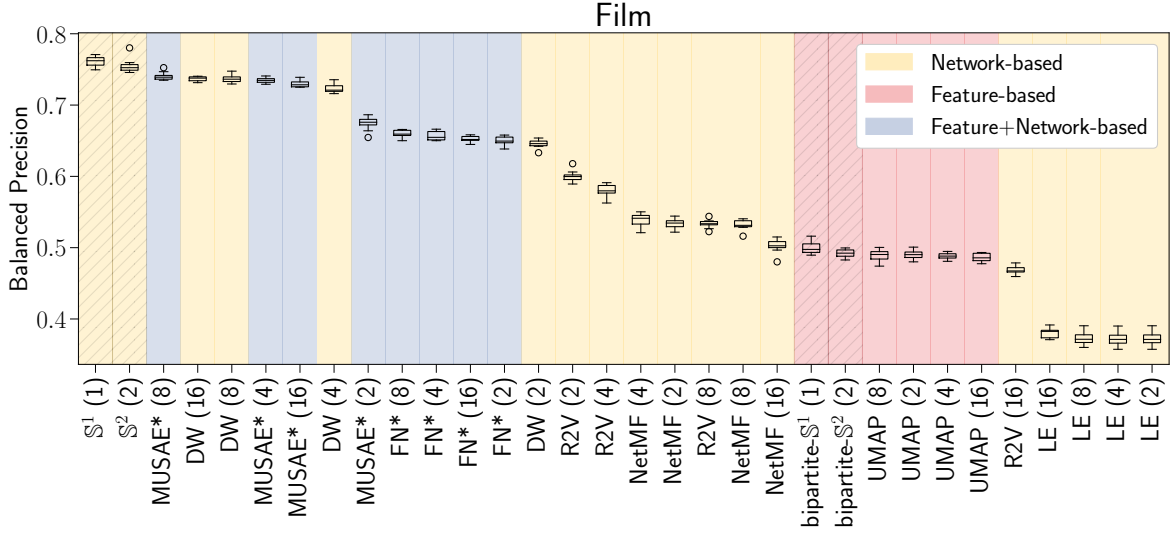

FIG. S31: Balanced precision of the distance-based link prediction task for the Film Dataset. The train/test sets are generated by randomly selecting a fraction $q = 0.1$ of existing links as positive samples, along with an equal number of randomly selected non-existing links as negative samples, ensuring a balanced test set. The remaining existing links form the training set. For balanced precision, assume the test set contains $L$ positive and $L$ negative links. The links are then sorted in ascending order based on their similarity scores, which are defined as the inverse of the hyperbolic distance between node pairs in $\mathbb{S}^1$, $\mathbb{S}^2$, bipartite-$\mathbb{S}^1$, and bipartite-$\mathbb{S}^2$, and as the inverse of the Euclidean distance between node pairs for the other methods. The balanced precision is then computed as the proportion of true positive links among the top $L$ ranked predictions. The results are averaged over 10 different splits. Our methods are highlighted with diagonal hatches. The abbreviations of the algorithms are as follows: DW – DeepWalk, R2V – Role2Vec, LE – Laplacian Eigenmaps, FN – FeatherNode. The numeric value in brackets indicates the embedding dimension. All other parameters are set to their default values.

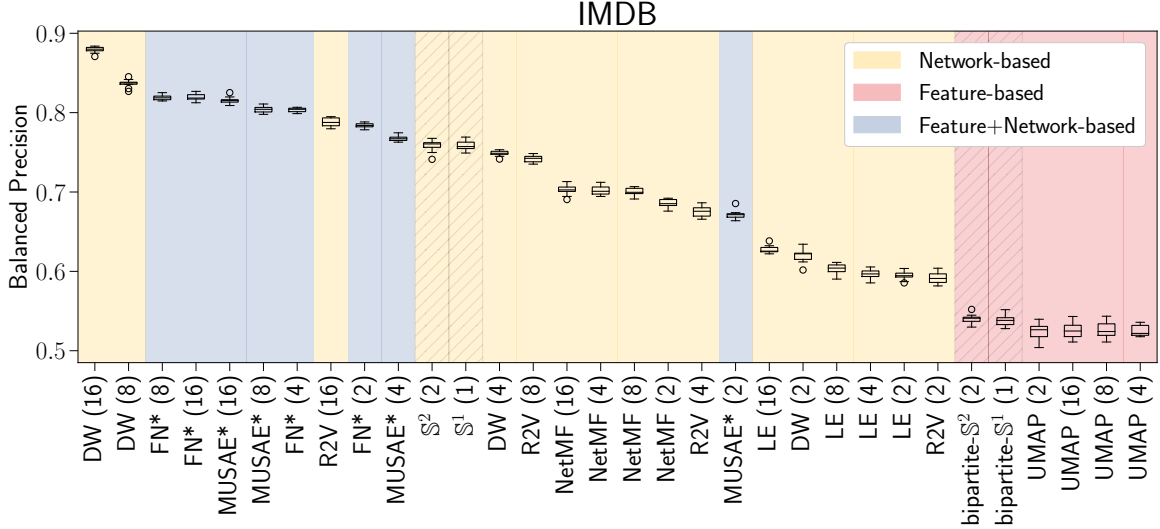

FIG. S32: Balanced precision of the distance-based link prediction task for IMDB dataset. See caption in Fig. S31 for more details.

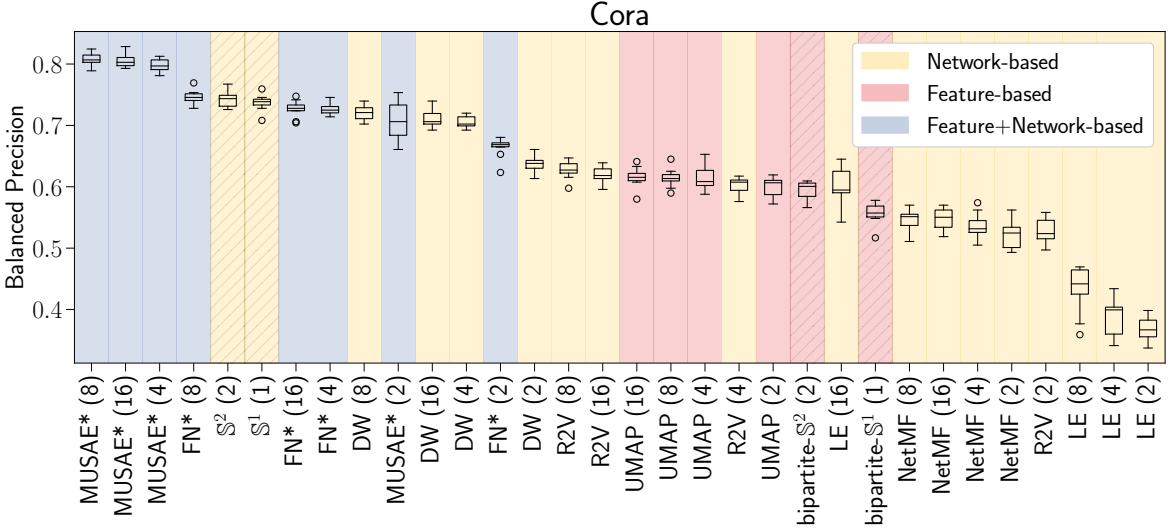

FIG. S33: Balanced precision of the distance-based link prediction task for Cora dataset. See caption in Fig. S31 for more details.

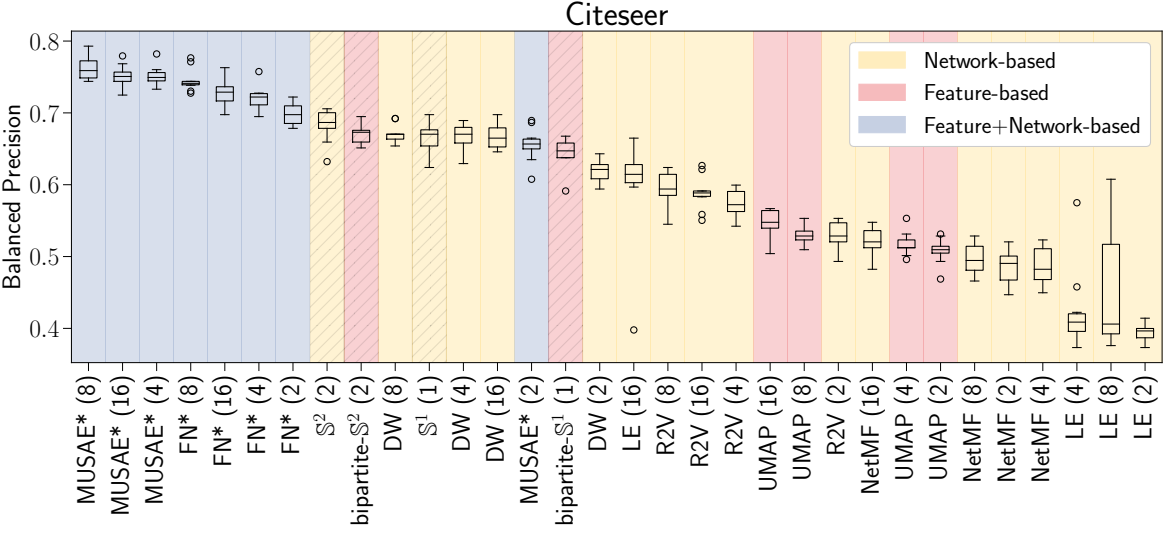

FIG. S34: Balanced precision of the distance-based link prediction task for Citeseer dataset. See caption in Fig. S31 for more details.

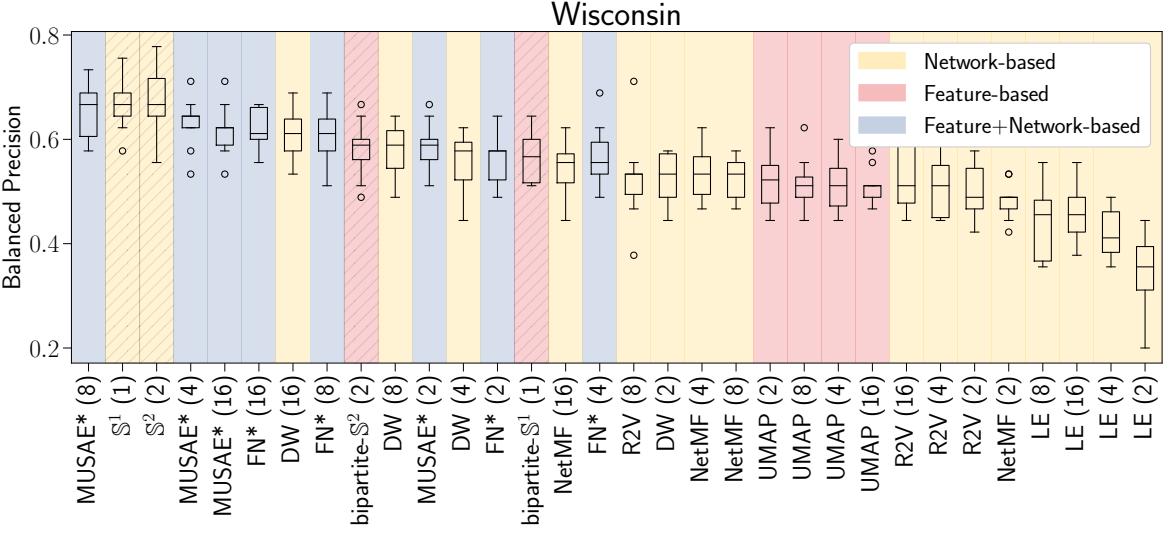

FIG. S35: Balanced precision of the distance-based link prediction task for Wisconsin dataset. See caption in Fig. S31 for more details.

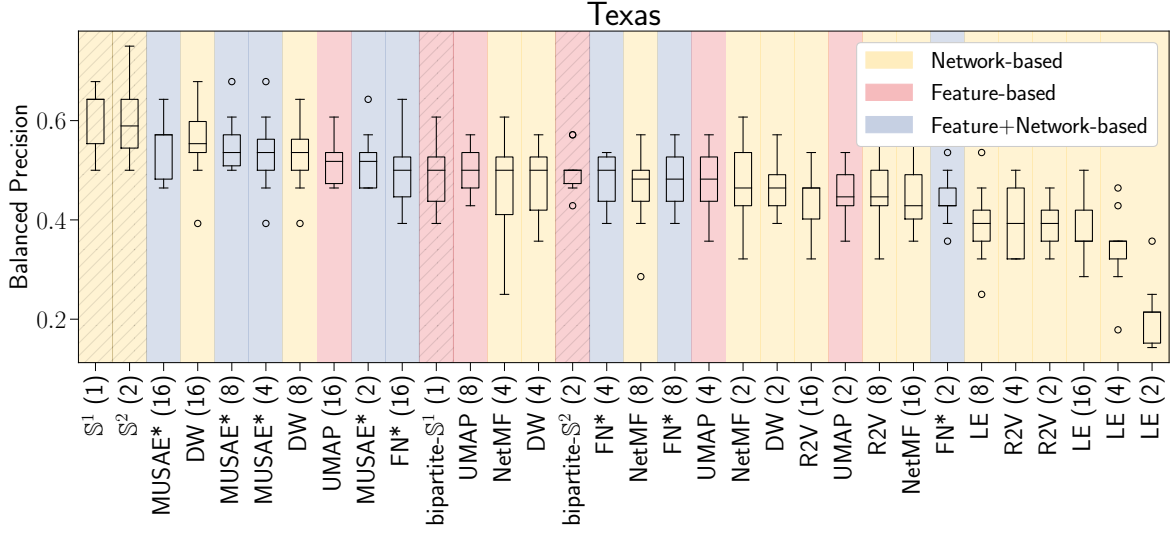

FIG. S36: Balanced precision of the distance-based link prediction task for Texas dataset. See caption in Fig. S31 for more details.

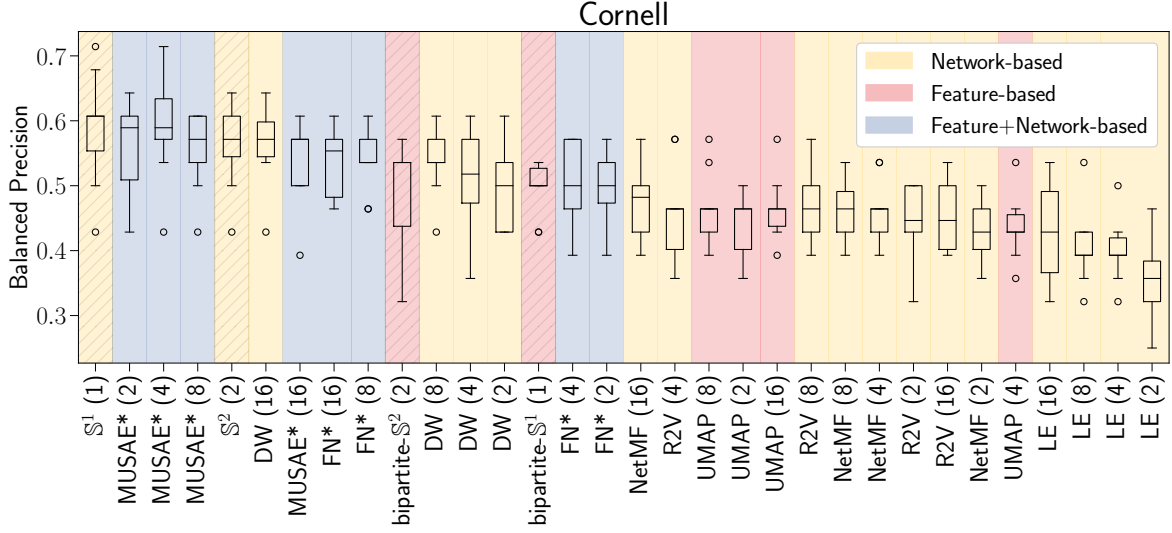

FIG. S37: Balanced precision of the distance-based link prediction task for Cornell dataset. See caption in Fig. S31 for more details.

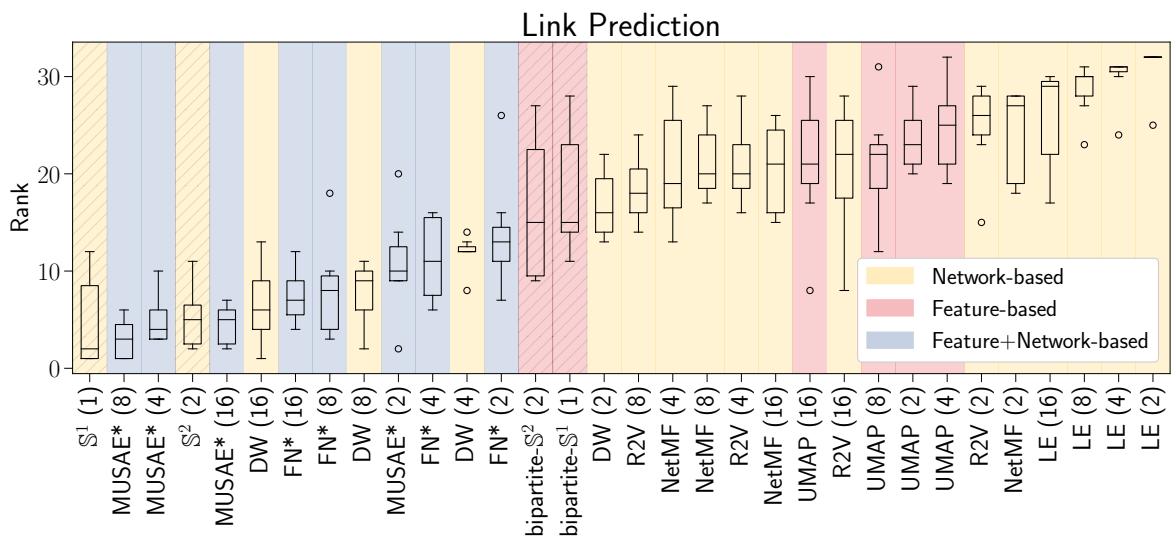

FIG. S38: The rank by balanced precision of network embedding methods across all datasets for the distance-based link prediction task.

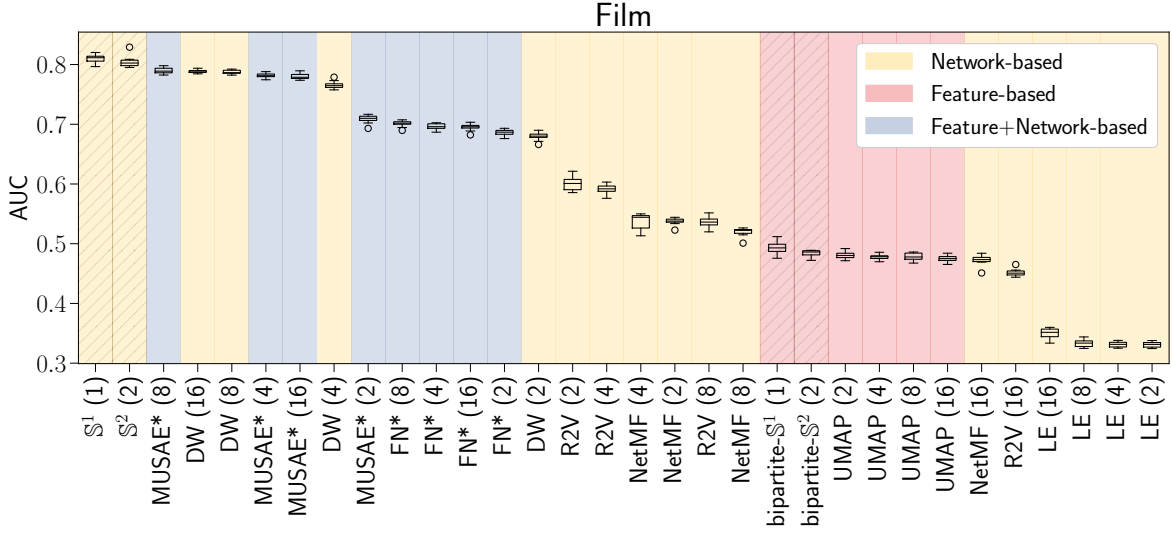

FIG. S39: AUC of the distance-based link prediction task for the Film Dataset. The train/test sets are generated by randomly selecting a fraction $q = 0.1$ of existing links as positive samples, along with an equal number of randomly selected non-existing links as negative samples, ensuring a balanced test set. The remaining existing links are used to form the training set. Similarity scores between node pairs in the test set are computed as the inverse of the hyperbolic distance between node pairs in $\mathbb{S}^1$, $\mathbb{S}^2$, bipartite-$\mathbb{S}^1$, and bipartite-$\mathbb{S}^2$, and as the inverse of the Euclidean distance between node pairs for the other methods. The AUC is then calculated as the area under the receiver operating characteristic (ROC) curve, which plots the true positive rate against the false positive rate at various similarity thresholds for the test set. The results are averaged over 10 different splits. Our methods are highlighted with diagonal hatches. The abbreviations of the algorithms are as follows: DW – DeepWalk, R2V – Role2Vec, LE – Laplacian Eigenmaps, FN – FeatherNode. The numeric value in brackets indicates the embedding dimension. All other parameters are set to their default values.

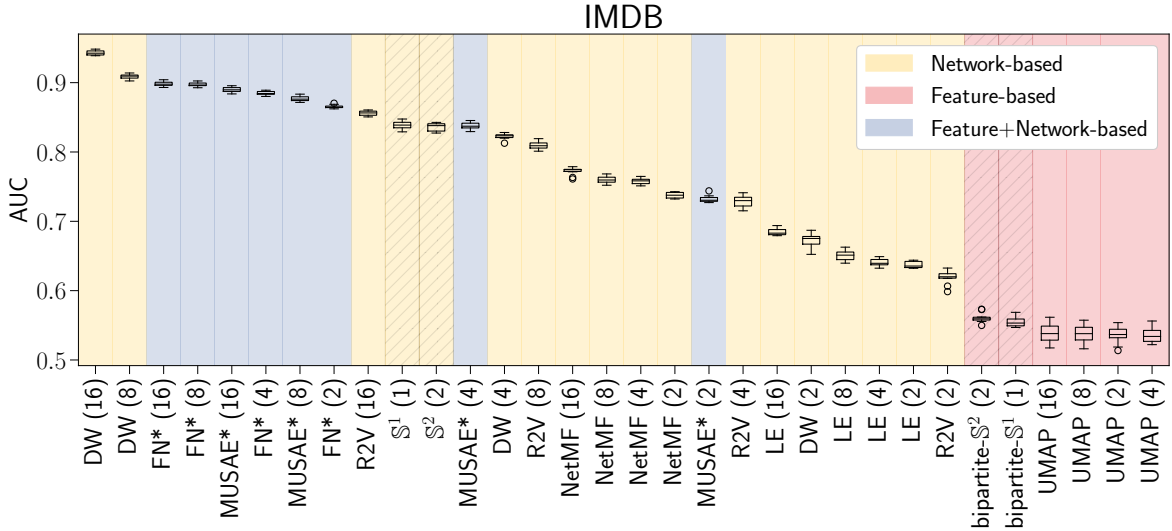

FIG. S40: AUC of the distance-based link prediction task for IMDB dataset. See caption in Fig. S39 for more details.

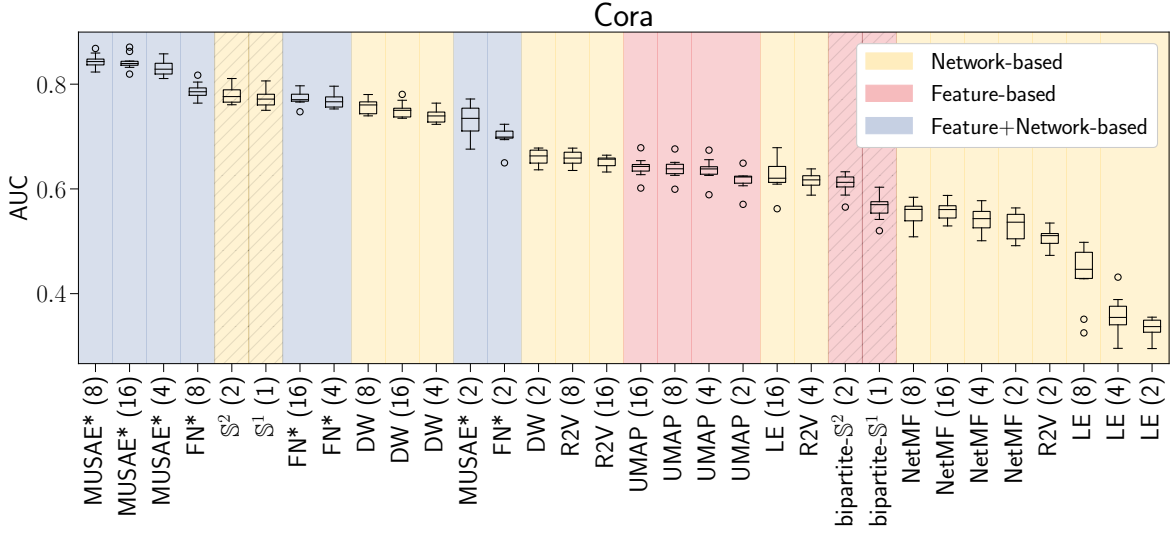

FIG. S41: AUC of the distance-based link prediction task for Cora dataset. See caption in Fig. S39 for more details.

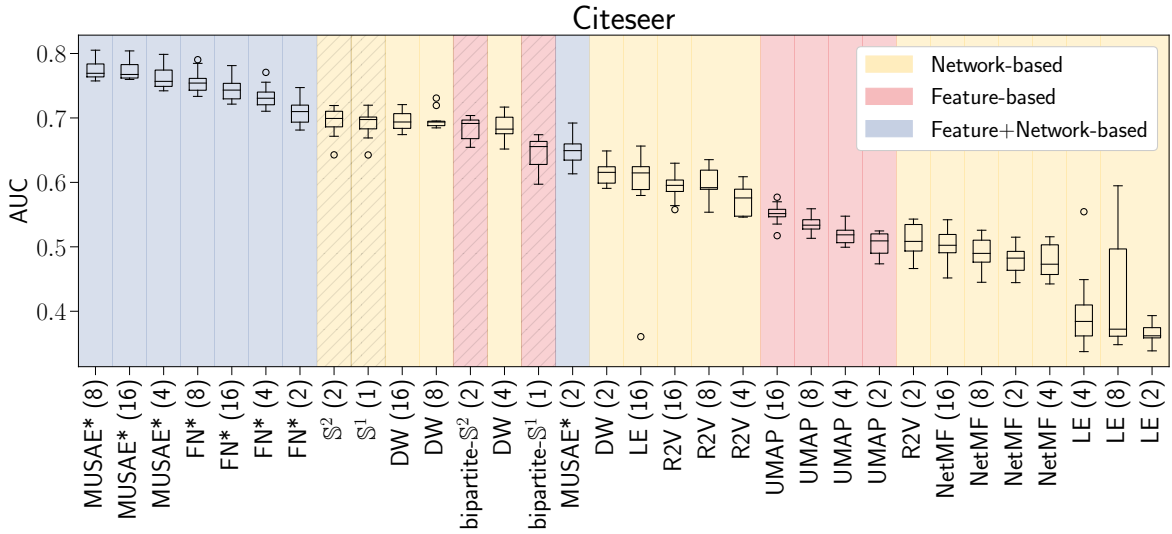

FIG. S42: AUC of the distance-based link prediction task for Citeseer dataset. See caption in Fig. S39 for more details.

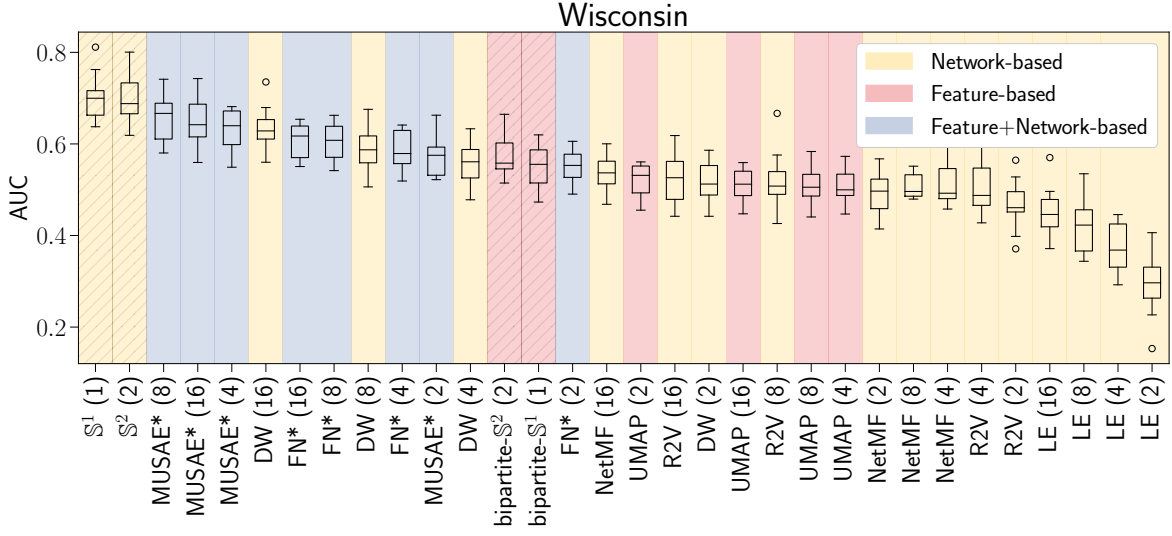

FIG. S43: AUC of the distance-based link prediction task for Wisconsin dataset. See caption in Fig. S39 for more details.

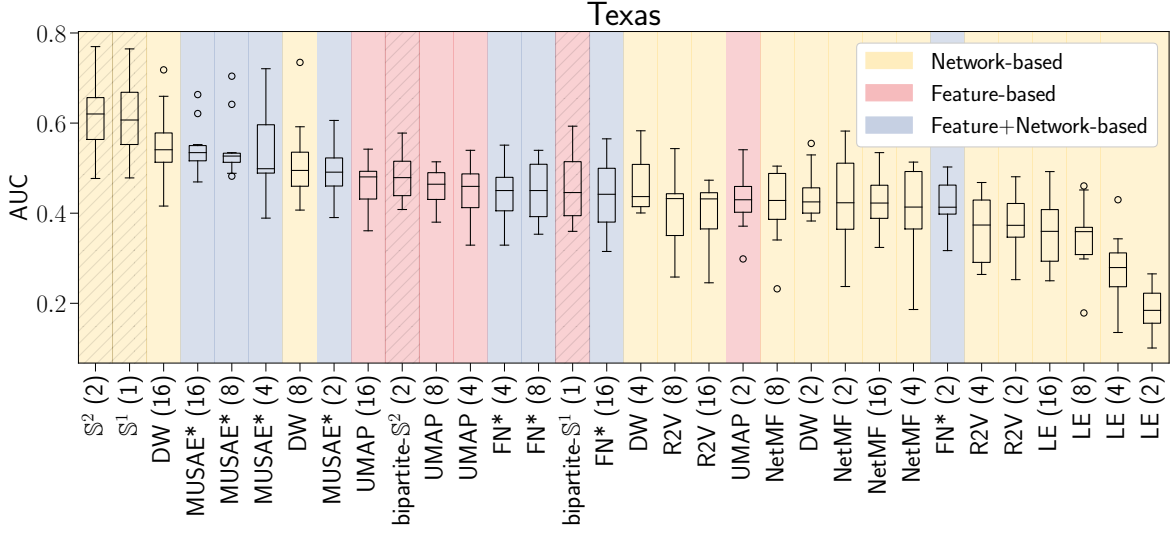

FIG. S44: AUC of the distance-based link prediction task for Texas dataset. See caption in Fig. S39 for more details.

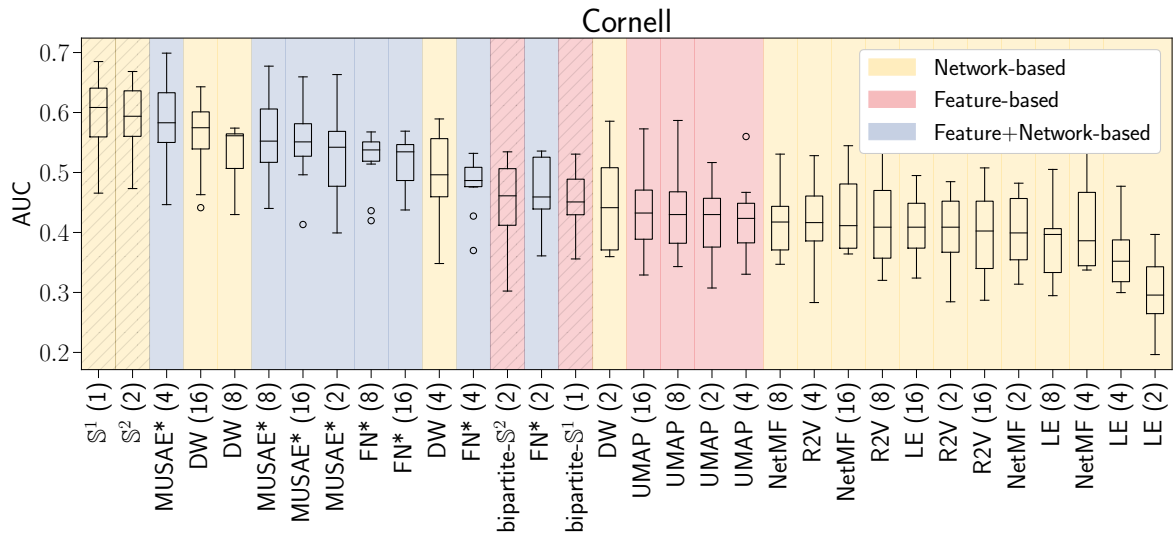

FIG. S45: AUC of the distance-based link prediction task for Cornell dataset. See caption in Fig. S39 for more details.

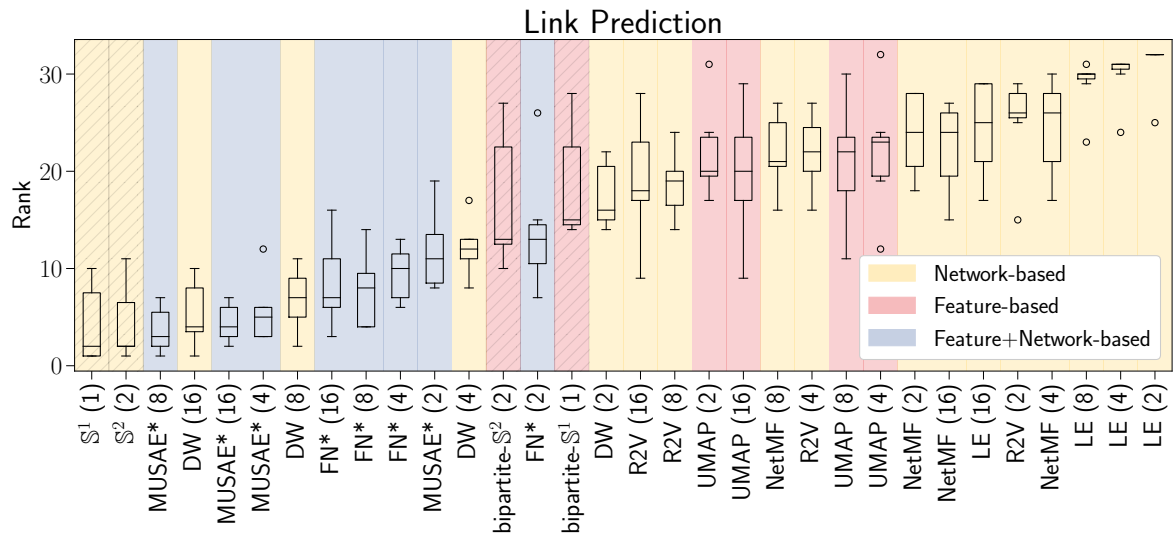

FIG. S46: The rank by AUC of network embedding methods across all datasets for the distance-based link prediction task.

## 11. VALIDATION OF THE TOPOLOGICAL PROPERTIES FOR THE MACHINE LEARNING DATASETS

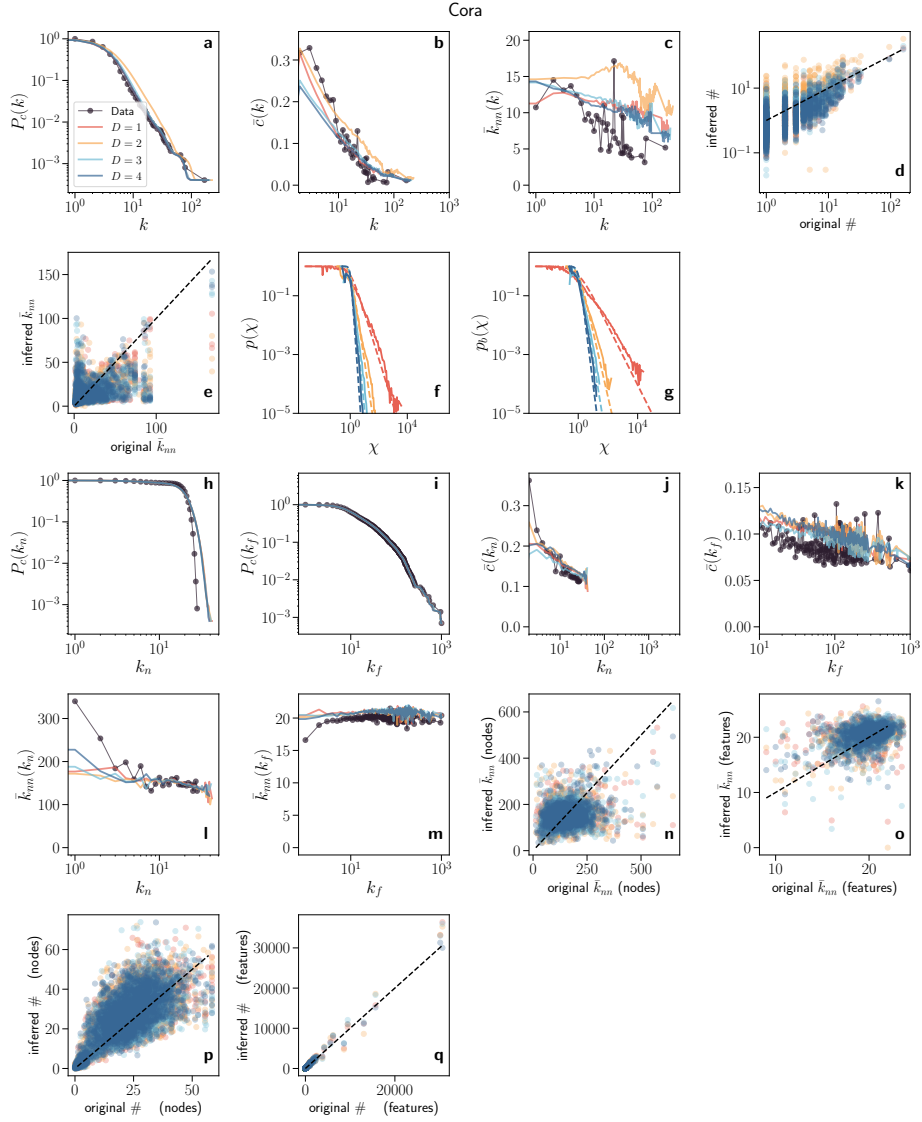

FIG. S47: Validation of the embeddings for Cora dataset. Plots (a-f) depict the topological properties of the unipartite networks. Meanwhile, plots (g-q) topological properties for the bipartite network, i.e., nodes' features. (a) Complementary cumulative degree distribution. (b) Clustering spectrum. (c) Average nearest neighbors degree in the function of degree. Scatter plots of the number of triangles (d) and the sum of degrees of their neighbors (e). (f, g) The expected connection probability is based on the inferred value of $\beta$ ($\beta_b$) (expected), and the actual connection probability is computed with the inferred hidden variables. Complementary cumulative degree distribution of (h) nodes and (i) features. Clustering spectrum of (j) nodes and (k) features. Average nearest neighbors degree of nodes (l) and features (m) in function of degree. Scatter plots of the sum of degrees of nodes (n) and features (o) of their neighbors and the number of triangles for nodes (p) and features (q). Symbols in (a-c,h-m) correspond to the value of these quantities in the original network, whereas the lines indicate an estimate of their expected values in the ensemble of random networks in a given dimension inferred by B-Mercator. This ensemble was sampled by generating 10 synthetic networks with the bipartite-$\mathbb{S}^D$ model and the inferred parameters and positions by B-Mercator. The error bars show the $2\sigma$ confidence interval around the expected value. The plots (d, e, n-q) show the estimated values of these two measures in the same ensemble of random networks considered above versus the corresponding values in the original network.

Citeseer

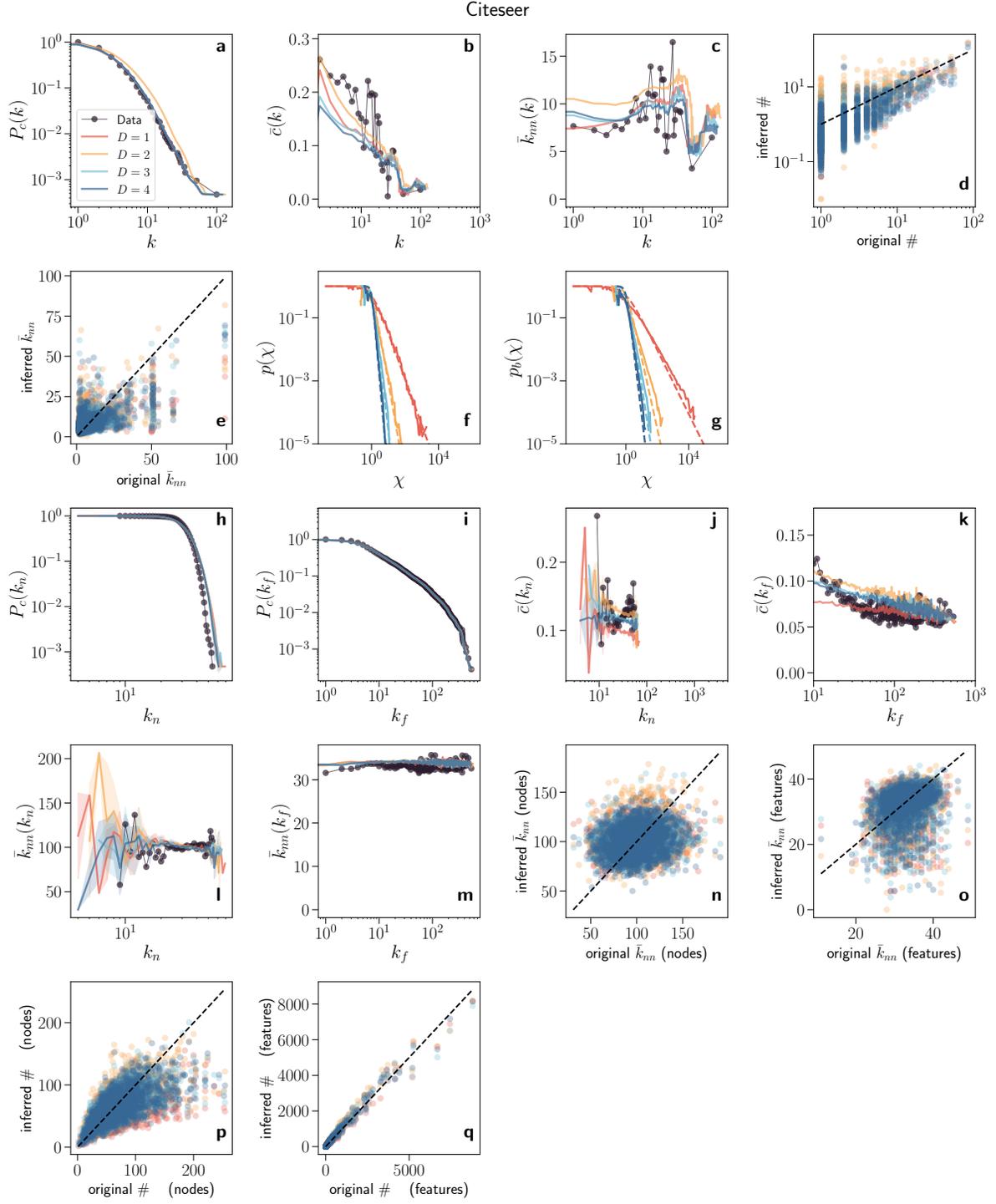

FIG. S48: Validation of the embedding for Citeseer dataset. See caption in Fig. S47 for more details.

Wisconsin

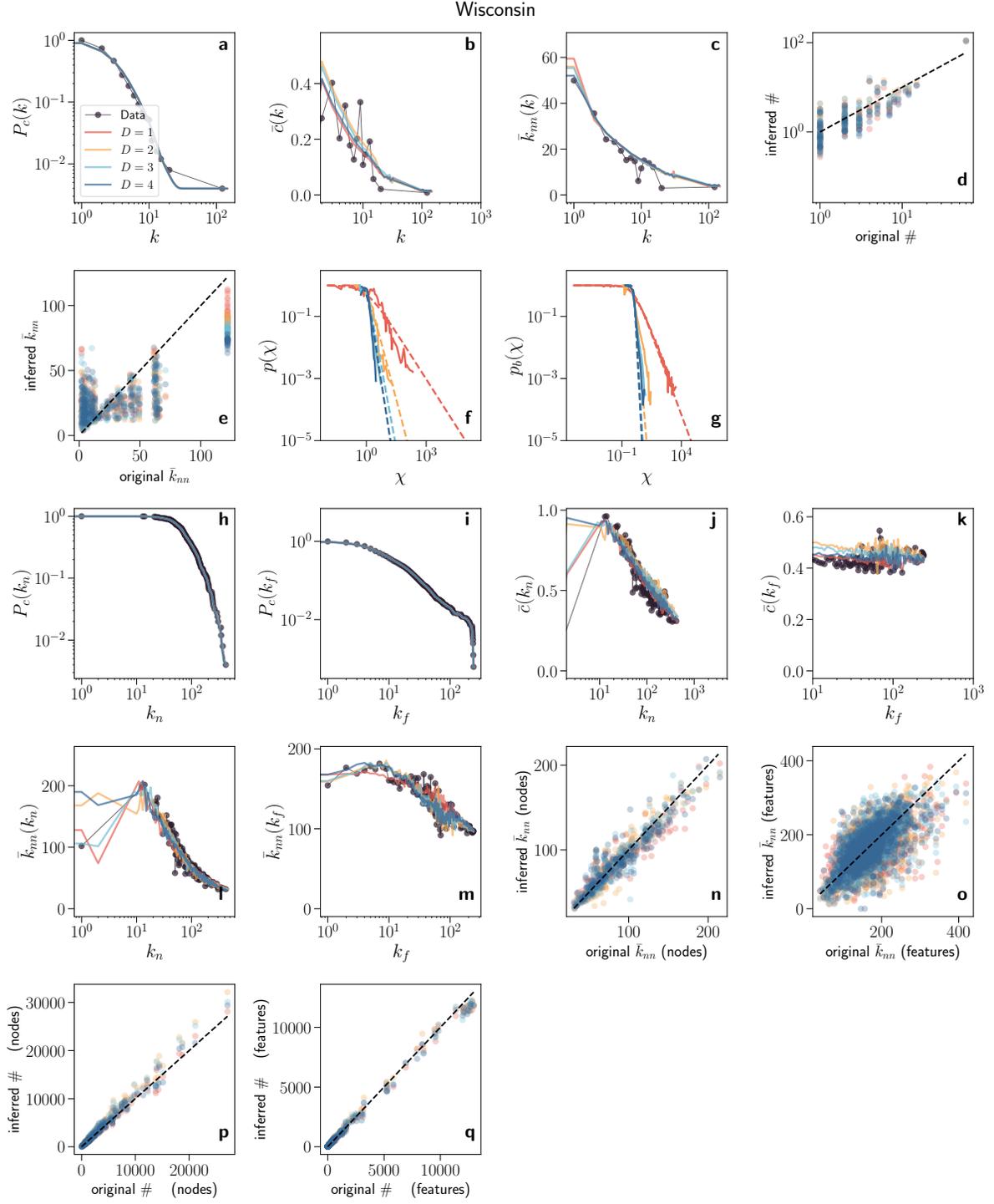

FIG. S49: Validation of the embedding for Wisconsin dataset. See caption in Fig. S47 for more details.

# SUPPLEMENTARY REFERENCES



\end{document}